%% file: paper.tex
\documentclass[12pt,headings=standardclasses]{scrartcl}

\usepackage{amsfonts}
\usepackage{amsmath}
\usepackage{amsthm}
\usepackage{subcaption}
\usepackage{amssymb}
\usepackage{adjustbox}
\usepackage{enumitem}
\usepackage[american]{babel}
\usepackage{csquotes}
\usepackage{enumitem}
\usepackage{mathtools}
\usepackage{longtable}

\DeclareMathOperator*{\argmin}{argmin}
\usepackage{centernot}
\usepackage{microtype}
\usepackage{appendix}

\usepackage{booktabs}
\usepackage{longtable}
\usepackage{array}
\usepackage{multirow}
\usepackage{wrapfig}
\usepackage{colortbl}
\usepackage{pdflscape}
\usepackage{tabu}
\usepackage{threeparttable}
\usepackage{threeparttablex}
\usepackage[normalem]{ulem}
\usepackage{makecell} 
\usepackage{xcolor}
\usepackage{siunitx}
\usepackage{hhline}
\allowdisplaybreaks

\usepackage[
  backend=biber,
  style = apa,
  maxcitenames=2,
  url=false,
  doi=true,
  dashed=false 
]{biblatex}
\AtEveryBibitem{%
  \clearfield{note}%
  \clearfield{urldate}%
  \clearfield{accessed}%
  \clearfield{issn}
}
\DeclareFieldFormat{apacase}{#1}
\AtEveryCite{\color{blue}}

\newtheorem{definition}{Definition}

\newtheorem*{theorem*}{Theorem}
\newtheorem*{definition*}{Definition}
\newtheorem*{proposition*}{Proposition}
\AtEndEnvironment{definition}{\noindent}
\AtEndEnvironment{assumption}{\noindent}
\AtEndEnvironment{proposition}{\noindent}
\AtEndEnvironment{theorem}{\noindent}

\usepackage{float}      
\usepackage[section]{placeins}
\usepackage{etoolbox}
\preto\subsection{\FloatBarrier}
\preto\subsubsection{\FloatBarrier}
\usepackage{tikz}
\usepackage{pgfplots}
\pgfplotsset{compat=1.18}

\usepackage{algorithm}
\usepackage{algpseudocode}

\usepackage{bm}

\addtokomafont{caption}{\footnotesize}
\setkomafont{captionlabel}{\sffamily\bfseries}

\usepackage[colorlinks=true, linkcolor=black, urlcolor=black, citecolor=blue]{hyperref}

\usepackage{cleveref}
\crefname{appendix}{Appendix}{Appendices}
\crefformat{appendix}{appendix~#2#1#3}
\crefname{algorithm}{algorithm}{algorithms.}
\Crefname{algorithm}{Algorithm}{Algorithms}

\usepackage{setspace}
\usepackage[a4paper, top=1in, bottom=1in, left=1in, right=1in]{geometry}

\usepackage{etoolbox}
\AtBeginEnvironment{footnotesize}{\setstretch{1.1}}

\title{Does Anxiety Improve Economic Decision-Making?}

\author{
    Ian Crawford\thanks{Ian Crawford \href{mailto:ian.crawford@economics.ox.ac.uk}{(ian.crawford@economics.ox.ac.uk)}: Department of Economics, University of Oxford, Manor Road, Oxford, OX1 3UQ, United Kingdom; Nuffield College, New Road, Oxford, OX1 1NF, United Kingdom. Thanks to Katerina Janezic for helpful discussions.}
    \and 
    Carl-Emil Pless\thanks{Carl-Emil Pless \href{mailto:cep@ifro.ku.dk}{(cep@ifro.ku.dk)}: Department of Food and Resource Economics, University of Copenhagen, Rolighedsvej 23, 1958, Frederiksberg C, Copenhagen, Denmark. Carl-Emil would like to thank participants in the PhD courses ``Behavioral Economics - Behind the Scenes'' and ``Food and Health Policy Analysis using Home-Scan Data''. Furthermore, Carl-Emil would like to thank Laura Mørch Andersen, Chen Zhen, Jette Bredahl Jacobsen, and Frank Jensen for many helpful discussions and thoughtful inputs. Finally, a special thanks to Sinne Smed, who originally developed the translated surveys our work relies on. A large part of this research was carried out while Carl-Emil Pless was visiting the Department of Economics, University of Oxford, Manor Road, Oxford, OX1 3UQ, United Kingdom, and he would like to thank the Department for their hospitality and support.}
    }

\begin{document}

\maketitle

\begin{abstract}
\linespread{1.1}
\noindent 
\small
We study the associations between everyday economic decision-making quality and people's emotional states. Using high-frequency, highly disaggregated consumer ``scanner'' data, we show that the cost of poor decision-making is substantial, on average equal to around half of day-to-day consumption budgets. While material circumstances help explain decision-making quality, how people \textit{feel} about those circumstances is equally important. Contrary to evidence that stress and worry impair performance in settings where distraction is costly, we find these same feelings are associated with improved decision-making for frequently made consumption choices. This is consistent with worry increasing attentiveness to decisions within households' locus of control. \\ \\
\noindent \textbf{Keywords:} revealed preference, scanner data, emotional states, self-control, stress, decision-making \\ \\ 
\noindent \textbf{JEL Codes:} D11, D12, D91
\end{abstract}

\newpage
\section{Introduction}
Effective economic decision-making or decision execution is central to economic welfare and living standards. We mean this not in the paternalistic sense that individuals may make decisions which a well-informed policy-maker or economist might regard as unwise or sub-optimal,\footnote{For example, \textcite{nardi_savings_2016} or \textcite{heiss_inattention_2021}.} but in the sense that they may fail to make good choices in terms of simply following their own preferences, and that by doing so, they leave money on the table.

Using high-frequency, highly disaggregated household scanner data, we show that the cost of poor decision-making in everyday consumption is substantial. On average, households appear to waste almost half their day-to-day consumption budget. The implication is economically significant: improving everyday decision-making so that purchases better reflect underlying preferences would yield a meaningful increase in living standards. This raises a question. If so much is at stake, why does the quality of decision-making vary so widely?

We examine correlates of decision-making quality in terms of households' objective, material circumstances. This complements similar exercises in, for example, \textcite{choi_who_2014}, \textcite{echenique_money_2011}, and \textcite{dean_measuring_2016}.  However, we also go further. Recent work in economics and psychology has shown that \textit{subjective} perceptions of material circumstances affect how people reason and make choices \autocite{mani_poverty_2013,haushofer_psychology_2014}. Using a novel aspect of our survey data, we study this explicitly. We find that decision-makers' \textit{emotional} state is as important as their material state. In particular, consumers who report being stressed or anxious---irrespective of their actual financial situation---consistently make better economic choices than those in similar objective circumstances who report no substantial feelings of stress or anxiety. This difference is significant both statistically and economically.

Our findings also help refine recent causal evidence on financial concern and job performance. While \textcite{kaur_financial_2025} show that reducing financial worries increases productivity by alleviating distraction in manufacturing work, our results suggest a countervailing mechanism in everyday consumption. Worry may increase attentiveness to relative prices and trade-offs, thereby improving decision-making ability in daily purchasing decisions.

Specifically, we demonstrate that feelings of stress, anxiety, and concerns about household finances serve as important predictors of how effectively consumers navigate complex everyday choices. These findings are based exclusively on real-world observational data and do not depend on any assumptions regarding preferences or preference heterogeneity. We argue that if the choice inefficiencies we identify were merely artifacts of high-frequency data, independent of anything intrinsic to the individual, then they would not correlate systematically with people's psychological states---let alone do so consistently and in the same direction across a number of different ways of describing the association. Therefore, the relationships we find between economic decision-making quality and psychometrically well-founded assessments of stress, anxiety, and worry, to us, suggest that standard economic theory is capturing something real and meaningful about human behavior that goes beyond being a practical ``as-if" predictive framework.

We use data drawn from a comprehensive and nationally representative consumer panel dataset from Denmark. The data records daily purchases at the level of the stock-keeping unit (barcode). This allows us to analyze decision-making with real-world products and real-world granularity and frequency.  In addition to standard demographics, the data contains a wide range of psychometrically well-founded measures related to respondents' self-control, stress, financial management, and planning.\footnote{For example, the data contains records of measures of self-control and stress using Tangney's Self-Control Scale \autocite{tangneyHighSelfControlPredicts2004} and Cohen's Perceived Stress Scale \autocite{cohen_global_1983}, respectively.}

The paper is structured as follows. In \cref{sec:literature}, we briefly review the related literature and situate our contribution within it. We describe the data used in \cref{sec:data}. Meanwhile, \cref{sec:concepts} is concerned with detailing how we intend to measure decision-making quality. We present our results in \cref{sec:results}, and, finally, \cref{sec:conclusions} discusses the implications of our findings and concludes the paper.

\section{Literature and Contributions}\label{sec:literature}
Decision-making ability is a topic of continuous interest in economics, as it speaks to a fundamental aspect of the subject that has potentially far-reaching policy implications. Therefore, several empirical studies have sought to identify how individuals' objective characteristics are associated with certain economic behaviors. For instance, \textcite{handel_socioeconomic_2024} shows that individuals with longer educations and higher incomes make different health insurance deductible choices than their less-educated or lower-income peers. Meanwhile, \textcite{jordan_role_2018} find that traits such as self-control and inquisitiveness are associated with more cooperative, pro-social and efficient behavior in experiments involving choices which affect others (dictator and third-party punishment games, etc.), whereas traits like leadership and caring are linked to more intuitive and, at times, less efficient behaviors (e.g. those may maximize one's own payoff but which fail to maximize the joint payoff). Other work that characterizes decision-making quality in terms of certain behaviors includes \autocite{epper_time_2020} and \autocite{de_bruijn_poverty_2022}, in which the ability to delay gratification is associated with wealth accumulation. Like these studies, we examine individuals' circumstances and decision-making.

There is a large literature in psychology and neuroscience on how emotions relate to economic decisions regarding time and risk (see, for example, the meta review and multinational analysis in \textcite{pertl_multinational_2024}). The general tenor of these findings is that individuals’ willingness to wait for delayed rewards or to take favorable risks is predicted by their self-reported incidental emotions \autocite{pertl_multinational_2024}. However, there is also substantial heterogeneity across studies, resulting in uncertainty about the average relationships between emotions and economic decisions in these areas.

The literature in economics is smaller and, as far as we know, entirely experimental in nature. \Textcite{ifcher_happiness_2011} finds a positive effect on time preferences---happier people tend to be more patient. \Textcite{campos-vazquez_role_2014} and \textcite{conte_risk_2018} both study the effects of sadness and anger on risk-taking behavior and find that sadder, angrier people are less risk averse---somewhat at odds with this, \textcite{conte_risk_2018} also investigates fearfulness and joyfulness, finding that these also increase risk-taking. \Textcite{cohn_evidence_2015} conducted an experiment on financial professionals, finding that participants who were primed to be in a fearful or worried state are more risk-averse. They point to this as a possible self-reinforcing mechanism for business cycles. Like these studies, we investigate emotional states. However, our focus is on the \textit{quality}, in the sense of efficiency, of economic decisions rather than the decisions themselves. Furthermore, we study observational data acquired from the real world, as opposed to a group of experimental subjects.

Our work also connects to a growing literature on the psychological effects of economic conditions (see \textcite{kaur_financial_2025} for a recent overview; also \textcite{haushofer_psychology_2014} and  \textcite{schilbach_psychological_2016}). This literature has highlighted that economic circumstances can shape cognition, preferences, and decision-making, with much of the evidence so far coming from settings characterized by financial adversity and measured either through well-being outcomes or through cognitive and preference-based tasks in psychometric or laboratory environments (e.g., \textcite{mani_poverty_2013, shah_scarcity_2015, carvalho_poverty_2016, fehr_poor_2022}). Building on these insights, \textcite{kaur_financial_2025} provide field evidence that alleviating financial concerns can improve work-related performance. We complement this agenda by examining everyday household consumption efficiency. Furthermore, whereas much of the existing literature is explicitly concerned with the adversity associated with poverty and financial scarcity, our work is not restricted to narrower areas of the income distribution. Thus, we provide observational evidence that subjective emotional states are an important correlate of decision-making ability even among households in similar objective circumstances.

A related strand of research already takes households’ self-reported feelings seriously, but in a macroeconomic setting. Since the 1940s, measures of consumer sentiment have been widely used as summary statistics for how households perceive the state of the economy and their own finances. In that literature, sentiment is viewed as a reliable early indicator of the overall strength of the economy. This is because (i) it may be correlated with current economic conditions, and current conditions predict future conditions, and (ii) because consumers’ feelings about the macroeconomy and their own personal financial situation may influence their spending habits, thus making sentiment a driving factor on the macroeconomy. In this tradition, \textcite{carroll_does_1994}, for example, uses the Michigan Index of Consumer Sentiment (ICS) to show that \textit{how consumers feel} is a good predictor of future consumption growth. Closer to our paper, \textcite{souleles_expectations_2004} uses the microdata from the Michigan Survey of Consumer Attitudes and Behavior (which is used to construct the ICS) to examine the rationality of households' expectations (in the macroeconomic sense) regarding inflation and other macroeconomic variables. They find that consumers' sentiment provides useful information on forecast errors regarding these variables. \textcite{acemoglu_consumer_1994} used a similar poll by Gallup in Britain also to test whether consumer confidence is consistent with the rational expectations hypothesis. They find that the predictive ability of confidence indicators is consistent with forward-looking behavior and that such data are useful predictors. We also study the relationship between sentiments and rationality. However, whereas this macroeconomic literature focuses on the model-consistency of beliefs about the macroeconomy, we focus on the quality of microeconomic decision-making revealed by households’ everyday consumption choices.

\textcite{choi_consistency_2007} takes a more experimental approach to measuring poor-quality decisions, and does so in an explicit utility-maximizing framework. They argue that previous studies struggle with the distinction between behaviors that are inefficient or irrational and those that merely appear so to a partially informed observer. They also argue that such studies are necessarily context-dependent and lack a straightforward, economically interpretable, and portable measure of decision-making quality. To accommodate this \textcite{choi_consistency_2007}, \textcite{choi_who_2014}, and \autocite{stango_we_2023} use incentivised experimental or quasi-experimental methods and an elegant, theoretically founded measure of decision-making quality based on fundamental ideas of revealed preference (RP).

These methods are particularly well-suited for identifying departures from efficient, rational decision-making because violations of the underlying axioms imply that the same choices could have yielded better outcomes subject to the same budget constraint. Put differently, they could achieve strictly preferred bundles of goods by reallocating expenditures and without increasing spending \autocite{varian_nonparametric_1982}. Thus, behavior that is irreconcilable with utility maximization is inefficient by definition: it is exactly equivalent to leaving money on the table \autocite{afriat_system_1973}. RP methods are also nonparametric in nature (that is, they do not rely on hard-to-verify assumptions about preferences) and allow for preference heterogeneity in an unrestricted way (by virtue of being applied at the level of the individual decision-maker). RP methods, therefore, are a powerful tool in diagnosing inefficiency, as they allow us to assess whether a consumer could have made a better choice while fully respecting their underlying preferences. We follow in the footsteps of \textcite{choi_consistency_2007} and \textcite{choi_who_2014} and adopt their nonparametric RP-based approach to examine maximizing behavior while allowing for unrestricted preference heterogeneity.

We also contribute to the literature that examines decision-making and maximizing behavior using scanner data, e.g., \textcite{echenique_money_2011} and \textcite{dean_measuring_2016}. However, we differ from these papers by using the data in their raw, disaggregated form. In contrast, these studies aggregate their scanner data over time (combining daily observations to the monthly level) and products (combining many SKUs into a few commodity groups) primarily to avoid the problem of zero demands. Highly disaggregated scanner data contain many zero demands. The difficulty with zero demands is that, when there is no transaction, there is no observed transaction price. This makes it impossible to establish revealed preference relations because we cannot value bundles at alternative prices.\footnote{We cannot, for example, calculate the cost of buying period $t$'s bundle at period $s$'s prices if the product was bought in period $s$ but was not in period $t$.} Aggregation deals with this by removing the zero demands, but it does so at the cost of strong auxiliary assumptions (principally regarding separability, but also through the choice of index number method). These assumptions muddy any ensuing results and make it difficult to determine whether the findings are driven by inefficiency in decision-making \textit{per se} or the auxiliary assumptions. In comparison, our measures of decision-making quality are relatively clean. Thus, we study choices at the same level at which they are actually made (individual items/SKUs) and can focus directly on departures from maximizing behavior.

\section{Data} \label{sec:data}
We use a household scanner panel data set called YouGov Shopper Denmark.\footnote{Until 2024, the panel was known as the ``Mini-Denmark'' panel and was maintained by GfK ConsumerScan Scandinavia, before GfK Panel Services was purchased by YouGov.} The purchasing behavior of each household is recorded by a designated respondent (termed the ``Shopping Responsible'') who scans the European Article Number (EAN) barcodes of all household purchases and reports the expenditure and the quantity purchased.\footnote{An EAN (European Article Number) code is a 13-digit barcode used worldwide to identify products. Canada and the United States use a 12-digit UPC code.} 

\subsection{Transactions}
Our raw data records 2,219,667 purchases in 2015.  We remove purchases that occur outside of Denmark, those where either the expenditure or the quantity purchased is missing, and those which YouGov have themselves classified as errors (for example, where the expenditure or quantity recorded is \emph{prima facie} absurd and clearly a typo by the respondent).\footnote{We add to these by excluding any purchases for which the observed expenditure is less than the smallest denomination of the Danish crown (0.5 DKK).} We measure purchases at the daily level. Thus, if a household purchases a certain EAN code more than once during the day (which is unusual but happens occasionally), we sum those transactions. Prices are measured per unit at the EAN code level (e.g., per item rather than per gram). These data-cleaning choices reduce the sample to 2,075,424 purchases by 2,400 households covering 49,068 products. 

In addition to the transaction data, respondents in the panel are asked to complete two surveys. The first captures standard sociodemographic characteristics, including household composition, income, the education of the primary earner, and so forth. The second asks questions about the respondents' behavioral traits, self-control, stress, planning behaviors, financial management skills, and general well-being. Not all panelists complete the entirety of these additional questionnaires. Therefore, the final sample we use for our analysis consists of 1,664 households. 

We provide a table of summary statistics for the transactions data in \cref{tab:transaction_stats}. On average, households shop on approximately 130 days per year, purchasing around 490 different products across roughly 970 transactions annually. Each shopping trip typically includes about eight products, with an average expenditure per trip of approximately 178 DKK.

\begin{table}[!htb]
    \centering
    \scriptsize
    \input{results/tables/descriptive/transactions_stats.tex}

    \caption{Descriptive statistics for the transactions data. The reported values are the means with the standard deviations reported in parentheses.}
    \label{tab:transaction_stats}
\end{table}

\subsection{Personality Traits}

\subsubsection{Disaggregated Traits}
The questionnaire on traits includes Danish translations of Tangney’s Self-Control Scale (SCS; \textcite{tangneyHighSelfControlPredicts2004}) and Cohen’s Perceived Stress Scale (PSS; \textcite{cohen_global_1983}), both widely validated psychometric measures of self-control and stress, respectively (see, e.g., \textcite{lee_review_2012}). Additionally, the survey includes statements that capture respondents’ attitudes toward financial management, well-being, and planning. All items are answered on a 5-point Likert scale ranging from 1 (“strongly disagree”) to 5 (“strongly agree”). Note that we shorten some of the variable names for simplicity in places. See \cref{tab:translations} for an overview. The survey was issued to participating panelists in November 2015.

Tangney’s SCS comprises 36 items related to habitualness, impulsivity, self-discipline, reliability, and procrastination. Self-control is particularly relevant in decision-making research as it involves the capacity to override dominant response tendencies and positively regulate behavior \autocite{de_ridder_taking_2012}. Higher self-control is consistently associated with desirable economic outcomes, including improved academic performance, fewer impulsivity problems \autocite{tangneyHighSelfControlPredicts2004}, increased savings rates, and reduced expenditure \autocite{baumeister_yielding_2002}. Conversely, low self-control correlates with detrimental economic behaviors, such as excessive consumption of tobacco, alcohol, and unhealthy foods \autocite{stautz_does_2018}.

Cohen’s PSS is one of the most frequently employed instruments to measure psychological stress, capturing whether individuals have felt overloaded, nervous, or out of control within the past month. While the original PSS consists of 14 items measured on a 4-point scale, we employ a psychometrically superior 10-item version using a 5-point Likert scale \autocite{lee_review_2012}. Stress has been shown to have significant implications for economic decision-making, potentially altering individuals’ preferences and consumption patterns. For example, stress has been linked to both under- and overeating, with chronic stress specifically increasing the preference for calorie-dense foods \autocite{torres_relationship_2007, oliver_perceived_1999}. Additionally, stress can trigger adaptive coping behaviors in consumption activities \autocite{moschis_stress_2007}, lead to impulsive spending, or, conversely, prompt overly cautious financial behaviors, such as avoiding spending altogether \autocite{durante_effect_2016}. 

We provide a comprehensive table of descriptive statistics for the two psychometric scales and the remaining survey questions we include in \cref{tab:survey_stats_all}.

\input{results/tables/descriptive/all_survey_stats_long.tex}

Both the PSS and SCS are widely used and validated psychological instruments with established track records in economics. Systematic reviews and meta-analyses consistently report strong internal consistency and test-retest reliability across different subjects \autocite{lee_review_2012,yilmaz_kogar_systematic_2024,de_ridder_taking_2012}. Importantly, both scales have also been used in economics research to study outcomes directly related to our setting. For instance, \textcite{haushofer_short-term_2016} use a 4-item version of the PSS in a large-scale cash transfer experiment, finding that transfers reduced perceived stress (alongside other improvements in psychological well-being).\footnote{In a similar fashion, \textcite{haushofer_economic_2020} find suggestive experimental evidence that health insurance reduces self-reported stress, as measured by the PSS.} Regarding self-control, \textcite{achtziger_debt_2015} administer the full SCS to a representative German sample and establish that low self-control predicts compulsive buying and debts. Similarly, using data from a large German household panel, \textcite{cobb-clark_predictive_2022} show that self-control, as measured using the Brief Self-Control Scale (a 13-item version of the original), predicts financial well-being, educational attainment, and labor market outcomes. Thus, the self-reported measures of stress and self-control we study are not only psychometrically sound but also relevant for understanding how certain psychological states relate to economic outcomes.

\subsubsection{Principal Components}
Because many survey items measure overlapping psychological and behavioral constructs, we summarize them via a principal components analysis (PCA) using polychoric correlations (suitable for ordinal Likert-scaled variables). Determining the optimal number of components involves balancing comprehensiveness, interpretability, and parsimony \autocite{velicer_construct_2000}. Although some of the original psychometric scales proposed different factor structures, e.g., five factors of self-control \autocite{tangneyHighSelfControlPredicts2004} and two for perceived stress \autocite{hewitt_perceived_1992}, the combined and additional statements may have significant overlap. Therefore, we combined various approaches to inform our decision on the final number of components to settle on. The combination of the ``eigenvalue greater than one'' criterion \autocite{kaiser_little_1974}, the scree plot slope \autocite{cattell_scree_1966}, and exploratory graph analysis using community detection \autocite{golino_investigating_2020} did not provide a unanimous answer to the number of components. Still, we settled on seven components, as this was the solution to two of the heuristics (the scree plot and exploratory graph analysis) and also provided clear interpretability. We summarize the extracted components in \cref{tab:summary_pca}.

\begin{table}[!htb]
    \centering
    \scriptsize
    \input{results/tables/pca/pca_summary.tex}
    \caption{Summary of the seven principal components. The table shows the component names based on content analysis, the number of items loading on each component, the proportion of variance explained, and internal consistency (Cronbach's alpha). Components represent key psychological and behavioral dimensions, including stress perception, self-regulation, conscientiousness, and financial attitudes.}
    \label{tab:summary_pca}
\end{table}

Items were assigned to components based on loadings exceeding 0.40, with each item assigned to the component where it loaded most strongly. The individual loadings and factors are available in full from \cref{tab:pca_loadings}. The identified components reflect distinct psychological and behavioral constructs: (1) \emph{Perceived Stress  \&  Anxiety}, capturing feelings of being overwhelmed and anxious; (2) \emph{Spontaneity and Disinhibition}, representing tendencies toward impulsive actions without deliberation; (3) \emph{Conscientiousness and Reliability}, reflecting self-discipline and dependability; (4) \emph{Poor Health Self-Regulation}, related to maintaining healthy habits and resisting temptation; (5) \emph{Financial Concerns}, encompassing attitudes toward debt, saving, and expenditure control; (6) \emph{Perceived Control}, indicating individuals’ sense of control over their lives; and (7) \emph{Shopping Impulsivity}, describing tendencies toward unplanned and impulsive purchases. These seven components explain approximately half of the total variance in the responses, confirming that the items measure several distinct yet interrelated dimensions. Internal consistency was acceptable for all components, except for \emph{Shopping Impulsivity}, which showed somewhat lower reliability but was retained due to its conceptual relevance.

\subsection{Sociodemographic Information}

Descriptive statistics for the sociodemographic variables are shown in \cref{tab:desc_stats}. Respondents are primarily women (77\%) in their mid-to-late 50s, aligning with the survey design wherein one household member (the ``Shopping Responsible'') reports purchases. Households typically consist of about two individuals, with 36\% being single-person households and around 30\% having children.

\begin{table}[!htb]
    \centering
    \scriptsize
    \input{results/tables/descriptive/desc_stats.tex}

    \caption{Overview of participating households and key descriptive statistics. The reported values are the means with the standard deviations reported in parentheses.}
    \label{tab:desc_stats}
\end{table}

Most respondents have vocational or medium-length educations, and the predominant labor market statuses are full-time employment (41\%) and retirement (30\%). Household incomes are concentrated in the lower to middle brackets, with roughly two-thirds earning below 500,000 DKK annually and only 6\% in the highest income category (above 800,000 DKK). These characteristics indicate that the sample is broadly representative of the Danish population according to national averages \autocite{statistics_denmark_statistical_2015}.

\section{Measuring Decision-Making Quality}\label{sec:concepts}

To measure decision-making quality, as opposed to measuring decisions themselves, requires a model of what constitutes optimal behavior. We follow  \textcite{choi_consistency_2007} and \textcite{choi_who_2014} by choosing the benchmark model:

\[ \underset{\mathbf{x}\in \mathbb{R}^{K}_{+}}{\text{max }}\;\;u(\mathbf{x})\text{ subject to } \mathbf{p}_{t}\cdot \mathbf{x} \le w_{t}.\]
Choices are made daily (indexed by $t$), the endogenous variable is the daily demand vector $\mathbf{x}_{t}$ for $K$ SKUs, the individual is a price-taker, and the daily budget $w_t$ is determined outside of the model. This model is consistent with any dynamic choice model in which preferences are intertemporally separable.

\subsection{Decision Quality}
Consider a single decision-maker and suppose that we have $T$ observations in a data set $\mathcal{D} = \left\{\mathbf{p}_{t}, \mathbf{x}_{t}\right\}_{t\in \{1,..,T\}}$.  For the moment, further suppose that the prices are fully observed.

Exact necessary and sufficient conditions for such data to be rationalizable by the standard model of consumer decision-making were given by \textcite{afriat_construction_1967} and made computationally tractable by \textcite{diewert_afriat_1973} and \textcite{varian_nonparametric_1982}.\footnote{\Textcite{afriat_construction_1967} provided necessary and sufficient conditions under which price-quantity observations can be rationalized by a well-behaved utility function, formalized as a non-cycling (cyclical consistency) condition. \Textcite{diewert_afriat_1973} subsequently demonstrated that Afriat's cyclical consistency was equivalent to checking the feasibility of an equivalent linear program, while \textcite{varian_nonparametric_1982} showed their equivalence to the Generalized Axiom of Revealed Preference (GARP).} The property that observational data must have in order to be consistent with perfect decision-making in this sense is described by the Generalised Axiom of Revealed Preference (GARP).\footnote{We say that $\mathbf{x}_{t}$ is directly revealed preferred to $\mathbf{x}_s$, written $\mathbf{x}_{t}R_{0}\mathbf{x}_s$, if $\mathbf{p}_{t}\cdot\mathbf{x}_{t}\ge\mathbf{p}_{t}\cdot\mathbf{x}_s$. We say that $\mathbf{x}_{t}$ is revealed preferred to $\mathbf{x}_s$, written $\mathbf{x}_{t}R\mathbf{x_s}$ if $\mathbf{p}_{t}\cdot\mathbf{x}_{t}\ge\mathbf{p}_{t}\cdot\mathbf{x}_{u}$, $\mathbf{p}_{u}\cdot\mathbf{x}_{u}\ge\mathbf{p}_{u}\cdot\mathbf{x}_{v}$,..., $\mathbf{p}_{v}\cdot\mathbf{x}_{v}\ge\mathbf{p}_{v}\cdot\mathbf{x}_s$ for some sequence of observations $\mathbf{x}_{t},\mathbf{x}_{u},\mathbf{x}_{v},...,\mathbf{x}_s$. In this case, we say that the relation $R$ is the transitive closure of the relation $R_{0}$.  We say that $\mathbf{x}_{t}$ is directly revealed strictly preferred to $\mathbf{x}_s$, written $\mathbf{x}_{t}P_{0}\mathbf{x}_s$, if $\mathbf{p}_{t}\cdot\mathbf{x}_{t}>\mathbf{p}_{t}\cdot\mathbf{x}_s$. }

\begin{definition}[GARP]
The Generalised Axiom of Revealed Preference: $\mathbf{x}_{t}R\mathbf{x}_{s} \implies$ not $\mathbf{x}_{s}P_{0}\mathbf{x}_{t}$. \end{definition}

GARP says that if one bundle is revealed to be preferred to another, then it should not be the case that the latter is also directly revealed to be strictly preferred to the former. Consistency of a panelist's behavior with GARP corresponds to perfect decision-making: their choices are \emph{exactly} consistent with the maximization of a stable, well-behaved (increasing, concave, continuous) utility function. Of course, the GARP-consistent consumer will not be doing this consciously, but nonetheless, their choices are as if they were doing so perfectly.

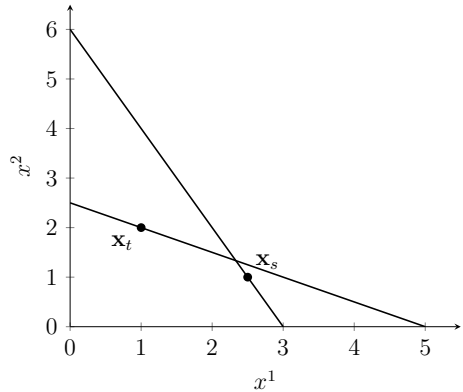
\begin{figure}[!htb]
    \centering

        \centering
        \begin{tikzpicture}[scale=0.75]
          \begin{axis}[
              axis x line=bottom,
              axis y line=left,
              xmin=0, xmax=5.5, 
              ymin=0, ymax=6.5,
              xlabel={$x^1$},
              ylabel={$x^2$},
              ytick={0, 1, 2, 3, 4, 5, 6},
              xtick={0, 1, 2, 3, 4, 5},
              ]
              \draw[thick] (axis cs:0,6) to (axis cs:3,0);
              \draw[thick] (axis cs:0,2.5) to (axis cs:5,0);
              \fill (axis cs:1,2) circle (2.2pt) node[below left] {$\mathbf{x}_t$};
              \fill (axis cs:2.5,1) circle (2.2pt) node[above right] {$\mathbf{x}_s$};
          \end{axis}
        \end{tikzpicture}
    \caption{A violation of rational choice.}
    \label{fig:violation}
\end{figure}

If the observation is that the decision-maker violates GARP, the question arises as to how it should be interpreted.  Consider the data set shown in \cref{fig:violation}. This is an example of prices and demands that cannot be reconciled with the standard rational choice model.

\begin{figure}[!htb]
    \centering
    \begin{subfigure}[b]{0.45\textwidth}
        \centering
        \begin{tikzpicture}[scale=0.75]
            \begin{axis}[
              axis x line=bottom,
              axis y line=left,
              xmin=0, xmax=5.5, 
              ymin=0, ymax=6.5,
              xlabel={$x^1$},
              ylabel={$x^2$},
              ytick={0, 1, 2, 3, 4, 5, 6},
              xtick={0, 1, 2, 3, 4, 5},
              ]
              \draw[thick] (axis cs:0,6) to (axis cs:3,0);
              \draw[thick] (axis cs:0,2.5) to (axis cs:5,0);
              \fill(axis cs:1,2) circle (2.2pt) node[below left] {$\mathbf{x}_t$};
              \fill (axis cs:2.5,1) circle (2.2pt) node[above right] {$\mathbf{x}_s$};
              \draw[very thick] (axis cs:0.25,5) to [bend right=12.5] (axis cs:1,2) to [bend right=25] (axis cs:5,.1);
              \draw[very thick] (axis cs:0.6,5) to [bend right=25] (axis cs:2.5,1) to [bend right=12.5] (axis cs:5,.45);
            \end{axis}
        \end{tikzpicture}
        \caption{Interpretation as inefficiency}\label{fig:inefficiency}
    \end{subfigure}
    \hfill
    \begin{subfigure}[b]{0.45\textwidth}
    \centering
    \begin{tikzpicture}[scale=0.75]
        \begin{axis}[
          axis x line=bottom,
          axis y line=left,
          xmin=0, xmax=5.5, 
          ymin=0, ymax=6.5,
          xlabel={$x^1$},
          ylabel={$x^2$},
          ytick={0, 1, 2, 3, 4, 5, 6},
          xtick={0, 1, 2, 3, 4, 5},
          ]
          \draw[thick] (axis cs:0,6) -- (axis cs:3,0);
          \draw[thick] (axis cs:0,2.5) -- (axis cs:5,0);
          
          \draw[very thick, domain=0.1:3, samples=100, smooth] 
            plot (axis cs:\x, {2*\x^(-0.25)});

          \draw[very thick, domain=2:4, samples=100, smooth] 
            plot (axis cs:\x, {97.65625/(\x^5)});
          
          \fill (axis cs:1,2) circle (2.2pt) node[below left] {$\mathbf{x}_t$};
          
          \fill (axis cs:2.5,1) circle (2.2pt) node[above right] {$\mathbf{x}_s$};
        \end{axis}
    \end{tikzpicture}
    \caption{Interpretation as preference change}\label{fig:instability}
    \end{subfigure}
    \caption{ \Cref{fig:inefficiency} shows the violation of rational choice in  \Cref{fig:violation}  as a failure to optimize efficiently; \Cref{fig:instability} shows it as a change in preferences.}
    \label{fig:both}
\end{figure}
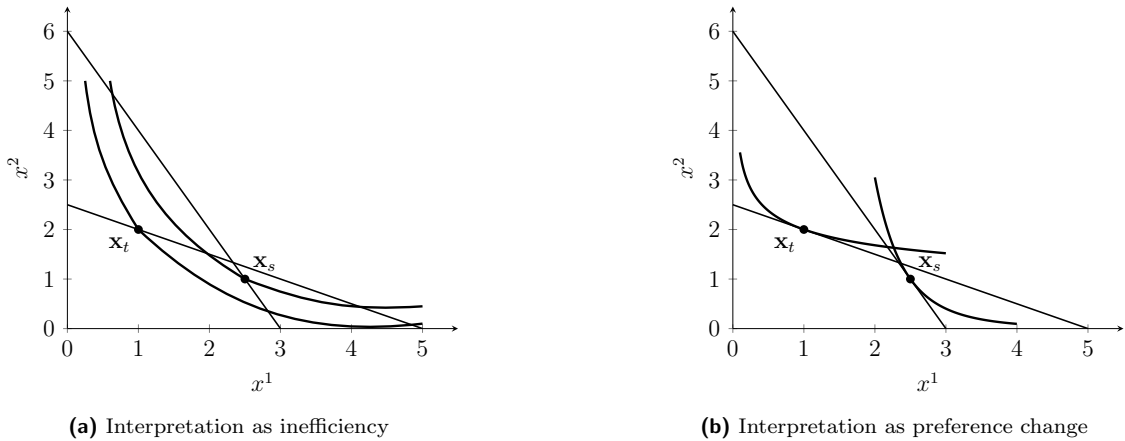

The next figure shows two possible interpretations of why that might be the case. \Cref{fig:inefficiency} shows that the decision-maker has well-behaved preferences but failed to optimize effectively. The consumer's behavior is said not to be \textit{cost-rationalizable} because there exist bundles that achieve at least as much utility as those chosen, but are less expensive. 

Note that cost-rationalizability does not restrict intertemporal decisions. It does not, for example, require consumers to buy bundles when they are at their cheapest. Using an example adapted from \textcite{polisson_rationalizability_2024}, suppose that $\mathbf{p}_1 = (2, 2)$ and $\mathbf{x}_1 = (2, 2)$, and  $\mathbf{p}_2 = (1, 1)$ and $\mathbf{x}_2 = (1, 1)$. This data set satisfies GARP and is cost-rationalizable, but the agent could clearly have saved money by buying the bigger bundle at the lower prices. This does not violate cost-rationalizability because the model does not restrict the decision-maker's choice of the timing of the utility targets. For instance, observation 1 could be spending on turkey and pumpkin pie in November during Thanksgiving, while observation 2 could be in early December. In this case, cost-rationalizability allows the decision-maker to derive higher utility from food during a festive period (even when prices are higher) and lower utility from food in normal times (even when prices are lower).\footnote{Similarly, a consumer purchasing paper towels on one day and soap on another is targeting different utility levels on each occasion; cost-rationalizability accommodates this without requiring that household inventories be modeled explicitly.} 

The alternative interpretation is shown in \Cref{fig:instability}. This illustrates the decision-maker as optimizing perfectly. However, their preferences change between the acts of choice. This consumer is said not to be \textit{preference rationalizable} because no utility function exists which assigns weakly higher utility to the chosen bundles than to any other bundle that is weakly cheaper at the prevailing prices. 

The notions of cost-rationalizability and preference-rationalizability are shown in \textcite{polisson_rationalizability_2024} to be definitionally equivalent under mild conditions and fully equivalent observationally---as both are characterized by GARP. However, the cost-rationalizability interpretation enjoys a considerable practical advantage when it comes to measuring departures from rational choice empirically. Whereas a description of rational choice violations in terms of preference change relies on some knowledge of preferences which are inherently \textit{unobservable}, an approach based on the cost-inefficiency description only relies on \textit{observables}. 

The idea---due to  \textcite{afriat_system_1973}---is that $\mathbf{x}_{t}$ is directly revealed preferred to $\mathbf{x}_s$ \emph{at efficiency level e}, written $\mathbf{x}_{t}R_{0}^{e}\mathbf{x}_s$, if $e\mathbf{p}_{t}\cdot\mathbf{x}_{t}\ge\mathbf{p}_{t}\cdot\mathbf{x}_s$.  Thus, expenditure on $\mathbf{x}_t$ at prices $\mathbf{p}_t$ needs, not just to exceed expenditure on $\mathbf{x}_s$, but to exceed it by a proportion of $(1-e)$ for $\mathbf{x}_t$ to be considered revealed preferred to $\mathbf{x}_s$. If $e$ is 0.95, for example, we only count bundles whose cost is less than 95\% of an observed choice as being revealed worse than that choice \autocite{varian_goodness--fit_1990}. It is as if the consumer simply threw away 5\% of their budget. The idea of revealed strictly preferred to \emph{at efficiency level e}  (denoted $P_{0}^{e}$) is defined analogously, but with a strict inequality. The definition of revealed preferred \emph{at efficiency level e}, written $\mathbf{x}_{t}R^{e}\mathbf{x}_s$, follows from the transitive closure of $R_{0}^{e}$. This, in turn, leads to a version of GARP that allows for imperfect decision-making.

\begin{definition}[GARPe]
The Generalised Axiom of Revealed Preference at efficiency level $e$: $\mathbf{x}_{t}R^{e}\mathbf{x}_{s} \implies$ not $\mathbf{x}_{s}P_{0}^{e}\mathbf{x}_{t}$. \end{definition}

The efficiency measure suggested in \textcite{afriat_system_1973} is the \emph{maximum} value of $e$ consistent with the data satisfying GARP$e$, which then provides our measure of decision-making ability.

\begin{definition}[Afriat Efficiency Index]
\begin{equation*}
    AEI \;=\; \sup \{\, e \;:\; \mathcal{D} \text{ satisfies GARP}_{e}\}\;\;\in[0,1].
\end{equation*}
\end{definition}

If $AEI=1$, then this equates to full efficiency; if $AEI=0.9$, then this equates to 90 percent efficiency. In addition to being simple to interpret in terms of the proportional cost of low-quality decision-making, AEI is easy to calculate and, because it is based on revealed preferences alone, entirely non-parametric in nature. Thus, the AEI enables the measurement of decision-making quality without requiring the specification of any parametric representation of preferences.

The Afriat Efficiency Index has become the most common way to measure how close observed choices are to perfectly rational and, hence, fully efficient decision-making.\footnote{For a more thorough overview, see, for example, \textcite{demuynck_revealed_2018}, \textcite{echenique_meaning_2022}, or \textcite{polisson_rationalizability_2024}, and \textcite{dziewulski_just-noticeable_2020} for a theoretical foundation in terms of being able to discern differences between bundles. See also \textcite{echenique_meaning_2022} for criticisms of this index.} Other noteworthy alternatives include the Houtman-Maks Index \autocite{houtman_determining_1985}, Varian's Index \autocite{varian_goodness--fit_1990}, the Money-Pump-Index \autocite{echenique_money_2011}, and the MASP Index \autocite{dean_measuring_2016}. While each of these methods has theoretical appeal, they are substantially more complex to compute and interpret. In fact, all except AEI are NP-hard\footnote{There is no known algorithm for computing them with solution times that are certain to only increase polynomially with the number of observations or, in the case of \textcite{dean_measuring_2016}'s MASP Index, in the number of revealed preference relations.}, making them infeasible to apply in our disaggregated setting. Moreover, the results of these different indices are known to correlate closely with AEI.\footnote{See \textcite{echenique_money_2011} and \textcite{dean_measuring_2016} for comparisons.}

\subsection{Dealing with Missing Transaction Prices}
We now return to the postponed issue of zero demands and missing prices. Missing prices make revealed preference analysis difficult because, if a product is purchased in period $s$ but not in period $t$, then the inequality  $\mathbf{p}_{t}\cdot\mathbf{x}_{t}\ge\mathbf{p}_{t}\cdot\mathbf{x}_s$ cannot be evaluated. This problem is pervasive in scanner data. 

We address this with an imputation/resampling approach. Specifically, for any item where demand is zero, we replace the missing price by drawing from the empirical distribution of observed prices for that item (independently, with replacement). With the missing prices sampled in this way, we compute the $AEI$. We then resample the missing prices and repeat the process. We use the average value of $AEI$ (denoted $\widehat{AEI}$) over these resamples to measure the household's decision-making ability. 

Resampling in this way accounts for the uncertainty involved in imputing prices in a reasonable and straightforward manner. We find that the expected $AEI$ converges rapidly in practice; fewer than $250$ iterations are typically sufficient to obtain a stable estimate. Nevertheless, we use $1,000$ resamples as a conservative default. A further advantage is that it works at the level of individual items. Importantly, this means that we need not worry about prices varying in package sizes.\footnote{For instance, it is implicitly assumed that a 0.5-liter bottle of soda should be a third of the price of the equivalent (but differently packaged) 1.5-liter soda if we were to use traditional price indices or aggregate vertically. See, for instance, \textcite{fox_non-linear_2014} for a discussion of how this affects statistical agencies' development of price indices.} Rather, this information is embedded in the unique code assigned to each product. In this way, we can accommodate the fact that preferences, as well as prices, may be tied to the packaging size of otherwise identical products in a (likely) non-linear fashion.

\section{Results}\label{sec:results} 
\subsection{The Distribution of Decision-Making Quality}
\begin{figure}[!htb]
    \centering
    \includegraphics[width=0.6\textwidth]{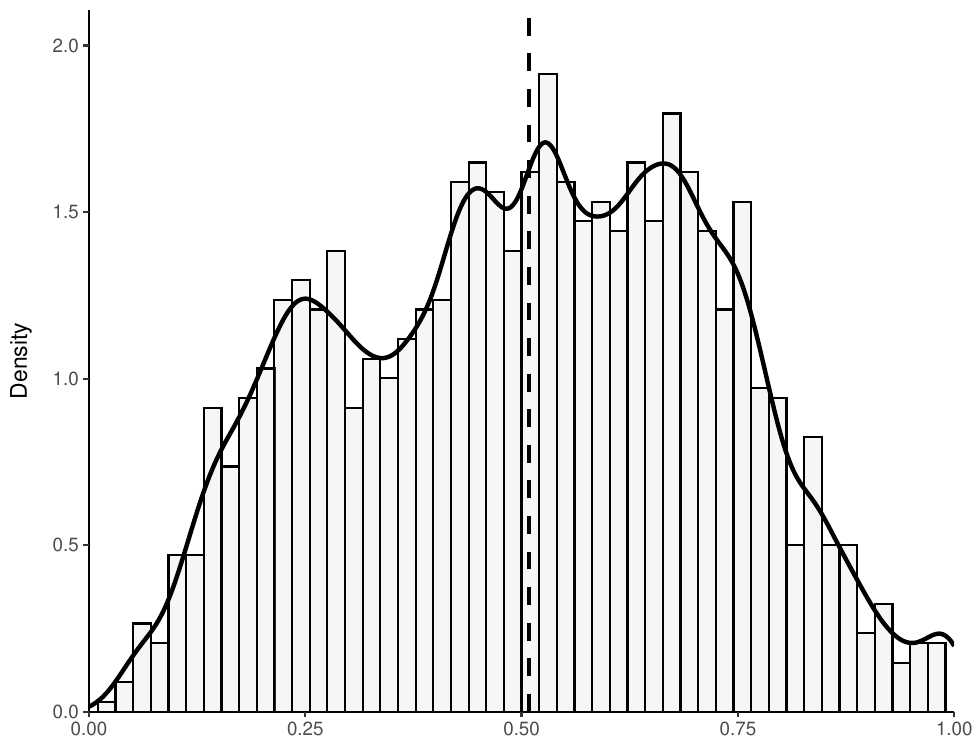}
    \footnotesize{\input{results/tables/descriptive/aei_stats}}
    \caption{The distribution of $\widehat{AEI}$. The vertical dashed line in the figure represents the mean.}
    \label{fig:dist_aei}
\end{figure}

The distribution of $\widehat{AEI}$ is illustrated in \cref{fig:dist_aei} along with some descriptive statistics.  Overall, the average efficiency of decision-making in these data is 0.5084. In other words, on average, these consumers' departures from full rationality and fully efficient decision-making come at the (substantial) cost of nearly half their budget. 

The measured decision-making quality in these data is lower than that suggested by earlier scanner-data studies. For instance, \textcite{echenique_money_2011} report an average efficiency of 0.97 using grocery data, while \textcite{dean_measuring_2016} find an even higher average of 0.99 within their food and beverage dataset. There are several possible explanations for the substantial difference.

First and foremost, our study includes, on average, five times as many observations per household compared to these two other studies. \Textcite{echenique_money_2011} and \textcite{dean_measuring_2016} aggregated their data to monthly observations (as discussed, largely to avoid the problem of missing transaction prices), and therefore, they only have 26 or 24 observations per household, respectively. In contrast, as we show in \cref{tab:transaction_stats}, we have an average of 129 daily observations per household. Because violations of the relevant revealed preference conditions are weakly increasing in the number of observations, daily data provides greater power to detect departures from perfectly rational decision-making. This is a methodological advantage, not a limitation: daily data reflects the level at which choices are actually made, whereas monthly aggregation may mask real inefficiencies by averaging them out. Our on-average five-fold increase in the number of observations per household thus likely accounts for the majority of the disparity in efficiency scores.

A second contributing factor could be the narrower subsets of commodities considered in those two studies. \Textcite{echenique_money_2011} examined 14 categories of groceries, and \textcite{dean_measuring_2016} restricted their attention to three food and beverage categories. 

From \textcite{varian_non-parametric_1983}, we know that if the demands for a subset of goods satisfy GARP, then they satisfy a necessary \textit{but not sufficient} condition for the separability of preferences with respect to that subset. A further necessary condition for separability is that the full set of goods (including the selected subset plus the outside goods) satisfies GARP as well. It is possible that the subsets of groceries analyzed in these studies looked close to rational, but that this further condition for separability was violated, albeit this was undetected because the relevant condition was not checked. By looking at choices over a much broader set of goods, we are therefore capturing this additional source of violations.

A third possible explanation has to do with the aggregation over products. Both referenced studies aggregate across SKUs/barcodes. We do not. This means that they have significantly fewer distinct goods than we do. We speculate that aggregation may diminish the sensitivity of revealed preference methods by masking violations of RP conditions, which occur at the individual product level. Furthermore, the fact that \textcite{dean_measuring_2016} assumes common prices and \textcite{echenique_money_2011} removes discounted prices due to partial observability may be removing the purchase-to-purchase price variability that is the source of inefficiency in the first place.\footnote{For instance, we document substantial variation in the prices of uniquely identified products, and our empirical procedure returns these varying prices according to the frequency at which we observe them being transacted.}

Whatever the reasons, it is clear that the consumers we examine are, on average, far less efficient decision-makers compared to previous studies based on scanner data.

\subsection{Correlates of Decision-Making Ability}
We present a number of regression results in \cref{tab:regressions} which relate decision-making quality to a variety of controls.  

In the second column of \cref{tab:regressions}, we report the results of a multiple regression using similar socioeconomic and demographic variables as \textcite{choi_who_2014}.  Older age groups (50–64 and 65+) appear to have significantly lower decision-making quality compared to the middle-aged reference group (35–49), while younger households are significantly more effective decision-makers. There is a consistent pattern of lower decision-making quality with increasing income.\footnote{The income groups 250,000DKK-500,000DKK and 800,000DKK+ have $p < 0.1$, while 500,000DKK-800,000DKK has $p < 0.05$.} The effects of education are insignificant. Having children in the household aged between 7-14 and 15-20 is associated with better decision-making, while having multiple people in a household\footnote{We do not exclude households with multiple people because the sample is only nationally representative at the household level. Furthermore, we would like our findings to be comparable to similar existing studies, e.g. \textcite{echenique_money_2011} and \textcite{dean_measuring_2016}, who do not reduce their sample to only include single-person households.} does not appear to matter significantly.\footnote{Notably, \textcite{dean_measuring_2016} find that multi-person households are more rational than single ones. Although our result does not indicate a statistically significant association, the negative sign appears to be the more intuitive finding. Especially when considering the known effects of preference aggregation, as \textcite{dean_measuring_2016} also note.} Finally, our results suggest that female heads-of-household tend to make less efficient decisions.

\begin{table}[htbp]
    \centering
    \scriptsize
    \input{results/tables/regressions/pca_regression_results.tex}
    \caption{Regressions of decision-making quality on sociodemographics and principal components. The standard errors are adjusted for the fact that the regressand is based on resampled imputed missing prices.}
    \label{tab:regressions}
\end{table}

Next, we turn to examining how the various self-reported behavioral statements correlate with decision-making ability. In the third column of \cref{tab:regressions}, we report the results of regressing decision-making quality on the seven principal components from \cref{tab:summary_pca} and the sociodemographics variables. For comparison, we also provide the reduced model that only includes the principal components in the second column. The analysis reveals two principal components that significantly predict decision-making quality (when controlling for sociodemographics): Self-Assessed Perceived Stress \& Anxiety and Financial Concerns. Both have positive associations with decision-making ability. 

Self-Assessed Perceived Stress \& Anxiety is characterized by high loadings on statements related to immediate emotional distress or feeling overwhelmed (See Table B.1 in Appendix B). In particular, high loading statements include ``Nervous and stressed?'', ``Difficulties piling up?'', and ``Cannot cope with all tasks?''. Higher scores on this component positively correlate with decision-making efficiency. Our interpretation is that respondents who self-report being stressed and anxious are, by their own admission, worriers. As such, they adopt a more deliberate approach to their consumption decisions, which may be aspects of everyday life they can exercise a degree of control over.

The Financial Concerns component also captures respondents' concerns and worries. However, in this case, the worries relate explicitly to their financial situation and personal financial management. Scoring highly on this component is consistent with strongly agreeing with statements such as ``Should save more'', ``Have more debt than should'', and ``Want better expenditure control''. The positive association of financial worries with decision-making ability suggests that these consumers recognize shortcomings or vulnerabilities in their current financial situation, which means they may be more attentive to their spending. Thus, this effect may result from two mechanisms. Either these respondents have more precarious financial situations that necessitate more careful spending, or they may premeditate future financial concerns by making more efficient decisions today.

In contrast, the components representing Spontaneity \& Disinhibition, Conscientiousness \& Reliability, Health Self-Regulation, Perceived Control, and Shopping Impulsivity are not statistically significantly related to efficiency in the fully controlled model. 

The fourth column of \cref{tab:regressions} combines the socio-economic variables and the psychometric components. The effect sizes of the two significant components decrease slightly. The attenuation of these effects in the full model suggests that their predictive power is correlated with background characteristics. For instance, it seems that certain health habits (as reflected by the Poor Health Self-Regulation component that is significant at the 10\% level in the PC-only model) are tied to certain preferences that are already captured by the sociodemographic characteristics. Furthermore, the income effects are no longer statistically significant in the full model. These variables may have served as a proxy for the financial concerns component, substantiating that worrying about finances is the more salient predictor of decision-making quality, rather than income itself. Nevertheless, the fact that the remaining sociodemographic effects persist highlights that \textit{both} objective sociodemographic factors \textit{and} the individual's own self-assessed perception of their circumstances independently contribute to heterogeneity in decision-making quality.

The profile of the efficient consumer that emerges from these results suggests that individuals who worry, whether broadly about circumstances in their everyday life or more specifically about finances, are more efficient in their consumption decisions. For instance, consumers who report financial stress may be more motivated to react to market conditions and relative price changes. Similarly, individuals who are generally stressed and anxious may be more thoughtful about domains they can control, such as how they allocate their consumption.

Given that the AEI is generally weakly decreasing in the number of observations, it is a compelling idea to include power controls, i.e., the number of observations and/or the number of products purchased by every consumer in the panel. However, this approach may be misguided, as these variables are inherently tied to both observable characteristics of the household (sociodemographics likely determine when and how much one shops) and unobservables, given that these data are self-reported. Therefore, including power controls introduces problems of endogeneity. Furthermore, we have found that they matter little for our overall findings.

\subsubsection{Unpacking the Principal Components}
While the principal components analysis allowed us to reduce the dimensionality of the 56 behavioral statements and summarize broad, latent constructs that are possibly relevant to economic decision-making, this approach implicitly assumes that the items comprising a given component are uniformly informative and that their predictive power is best captured through linear aggregation (see, for example, \textcite{artigue_principal_2019}). To relax these assumptions and allow for a more flexible, data-driven assessment of behavioral correlates of decision-making quality, we estimate three alternative models based on the penalized Least Absolute Shrinkage and Selection Operator (Lasso) regression \autocite{tibshirani_regression_1996} using all 56 original statements alongside the full set of sociodemographic controls. These are: the standard Lasso, Group Lasso, and Sparse Group Lasso.

Lasso allows us to directly identify which individual behavioral tendencies most strongly predict variation in economic decision-making quality. A key advantage of the Lasso is that it allows us to include behavioral statements that would otherwise be excluded in a PCA framework if they did not load strongly onto a single component. Finally, because the Lasso selects predictors based on their direct association with the outcome rather than relying on latent components that may or may not influence decision-making, it can provide more detailed insights into which specific variables explain the greatest share of variance in the quality of decision-making.\footnote{As \textcite{hadi_cautionary_1998} note, principal component regression may fail to explain any variation in $\widehat{AEI}$, even if it accounts for a large share (in our case, roughly 50\%) of the variance in the latent constructs.} 

We implement the Lasso estimator in its standard form by standardizing all predictors and selecting the regularization parameter ($\lambda$) using ten-fold cross-validation:
\begin{equation}\label{eq:lasso_estimator}
    \widehat{\boldsymbol{\delta}}^{\text{Lasso}} = \argmin_{\alpha, \; \boldsymbol{\delta}} \left\{ 
    \frac{1}{2N} \sum_{i=1}^N \left( \widehat{AEI}_i - \alpha - \mathbf{W}_i \boldsymbol{\delta} \right)^2 
    + \lambda \left\| \boldsymbol{\delta} \right\|_1 \right\}
\end{equation}
where  $\mathbf{W}_i$  is the matrix of sociodemographic and behavioral variables, the corresponding vector of coefficients is $\boldsymbol{\delta} $, and $\alpha$ is a constant.

A problem with the Lasso is that it ignores the thematic clustering that our initial principal component analysis revealed. Furthermore, many of the sociodemographic variables are grouped in ways that make it advantageous to consider their collective effects alongside their individual importance. To accommodate such groups, we also estimate a Group Lasso (GL) \autocite{yuan_model_2006} that applies an $\ell_2$-penalty across pre-specified groups of variables. Suppose we divide $\mathbf{W}_i$ into $g = 1,\dotsc, G$ non-overlapping groups such that each group $g$ indexes a subset of coefficients in $\boldsymbol{\delta}$, then we define the Group Lasso estimator as:
\begin{equation}\label{eq:group_lasso}
    \widehat{\boldsymbol{\delta}}^{\text{GL}} = \argmin_{\alpha, \; \boldsymbol{\delta}} \left\{
    \frac{1}{2N} \sum_{i=1}^N \left( \widehat{AEI}_i - \alpha - \mathbf{W}_i \boldsymbol{\delta} \right)^2
    + \lambda \sum_{g = 1}^G \left\| \boldsymbol{\delta}(g) \right\|_2 \right\}.
\end{equation}
This specification ensures that the individual coefficients within the group collectively shrink toward zero, rather than the standard Lasso, which picks up the most powerful signals and drops the rest\footnote{As \textcite{mullainathan_machine_2017} note using a house price example, in situations with many highly collinear predictors, the standard Lasso may arbitrarily pick a single predictor among the collinear variables and disregard the others (in their example, they show that over 10,000 iterations, half of their variables are unused in each run).}. The GL addresses this collinear instability by ensuring that individual variables within each pre-specified group are selected or excluded as a unit, rather than competing against each other in the variable selection process. This collective selection mechanism prevents the arbitrary choices among highly correlated predictors that \textcite{mullainathan_machine_2017} describe as akin to a ``flip of the coin'', where the Lasso may randomly favor one correlated variable over another across different model runs.

While the Lasso returns a sparse set of predictors and the GL returns a sparse set of groups, we are also interested in reducing the number of predictors within each group. To that end, we also estimate a Sparse Group Lasso (SGL) \autocite{simon_sparse-group_2013}. To achieve group-wise interpretability and within-group sparsity, we solve
\begin{equation}\label{eq:sgl}
    \begin{aligned}
        \widehat{\boldsymbol{\delta}}^{\text{SGL}} = \argmin_{\alpha, \; \boldsymbol{\delta}} \Bigg\{ \;
        & \frac{1}{2N} \sum_{i=1}^N \left( \widehat{AEI}_i - \alpha - \mathbf{W}_i \boldsymbol{\delta} \right)^2 \\
        & + \lambda \left[ (1 - \omega) \sum_{g = 1}^G \left\| \boldsymbol{\delta}(g) \right\|_2 + \omega \| \boldsymbol{\delta} \|_1 \right]
        \Bigg\},
    \end{aligned}
\end{equation}
where $\boldsymbol{\delta}{(g)}$ are the coefficients for group $g$, $|g|$ is the size of the group, and $\lambda$ is the regularization parameter we select using ten-fold cross-validation. The parameter $\omega \in [0, 1]$ controls the mixture of group vs. individual sparsity. If we set $\omega = 0$ we recover the pure GL, while setting $\omega = 1$ returns the standard Lasso. The innovation behind \cref{eq:sgl} is for values of $0 < \omega < 1$, which first discards entire groups if they lack predictive power, and then selects the most predictive individual items from within the retained groups.

Thus, by varying $\omega$, we move from an ``all-group'' selection to an ``all-item'' selection. Empirically, we set $\omega = 0.95$ by searching the grid $\omega \in [0.05, 0.95]$ in $0.05$ intervals for the value that minimizes the mean-squared error (and provides the best-performing model). In our context, the SGL has the advantage that it encourages the entire behavioral constructs or sociodemographic blocks to be included or excluded block-by-block, which resolves the instability that the pure Lasso exhibits under collinearity \autocite{mullainathan_machine_2017}. Furthermore, the individual penalty ensures that each retained group is reduced to its most predictive statements.

\begin{figure}[!htb]
    \centering
    \includegraphics[width=0.8\textwidth]{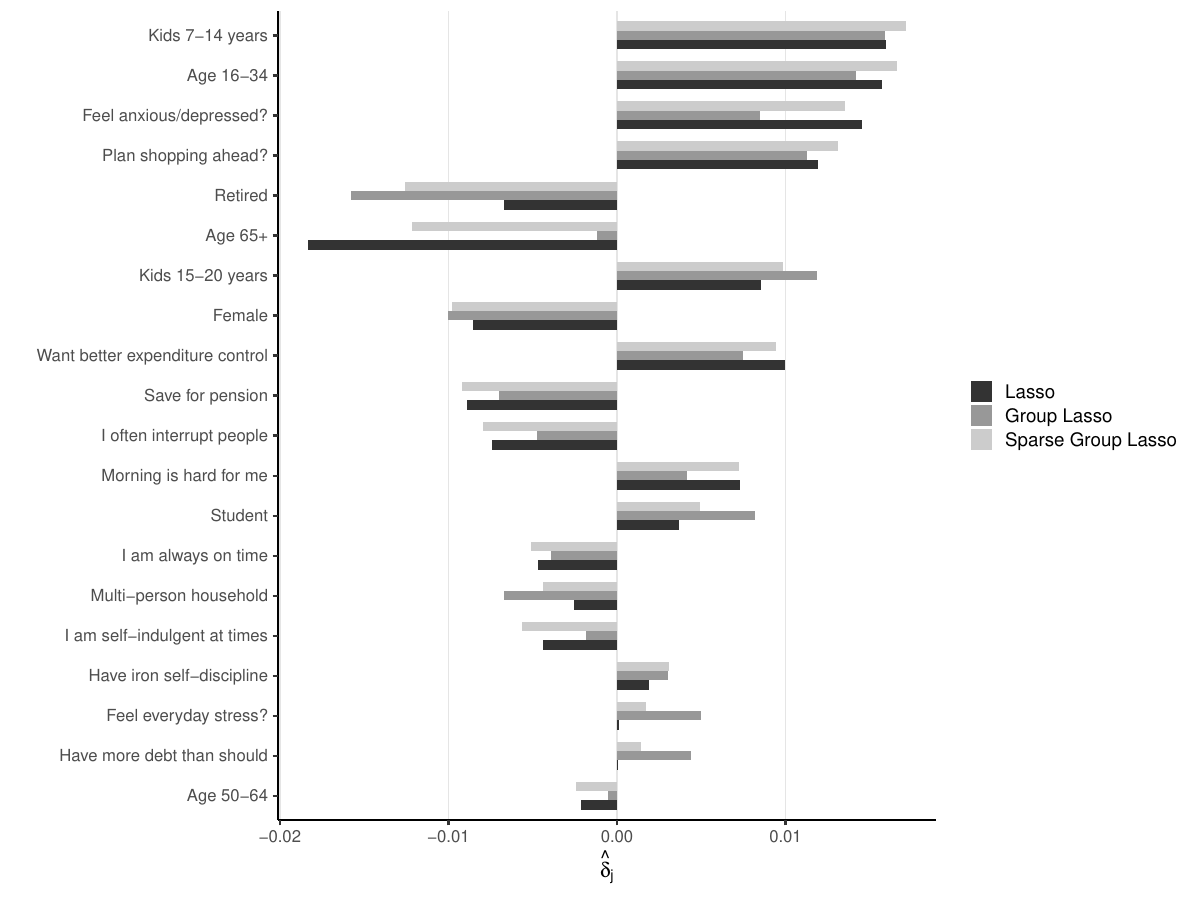}
    \scriptsize
    \input{results/tables/lasso/model_comparison.tex}
    \caption{Selected predictors for Lasso, Group Lasso, and the Sparse Group Lasso. The figure presents the 20 variables that are most frequently selected across models. Below the figure, we also provide a summary of model statistics for the three different models.}
    \label{fig:lasso_predictors}
\end{figure}

We provide an overview of the 20 best predictors from all three variations of the Lasso in \cref{fig:lasso_predictors} and refer to \cref{app:lasso} for the full models. Interestingly, the Lasso results retain and expand upon the core insights from the principal component regressions. The two components found to be statistically significant are echoed in the Lasso-based models through the selection of some of their highest loading items. In particular, all three models count ``Feel anxious/depressed?'' and ``Want better expenditure control'' among their most predictive variables. Interestingly, having saved for your pension shows up as an important negative predictor of decision-making ability, which may strengthen the interpretation that it is worry about your current financial situation, rather than your actual financial situation, that predicts high-quality economic decision-making.\footnote{This particular statement: ``I save/have saved for my pension'' loads negatively ($-0.37$) onto the financial concern component, but not cleanly enough to warrant inclusion. Therefore, its effect on decision-making is also inverted compared to the financial concern component.} 

Another pattern that emerges from \cref{fig:lasso_predictors} is that planning your purchases (strongly agreeing with the statement ``I plan shopping ahead?'') and exerting self-restraint (by scoring low on the ``I often interrupt people'' and ``I am self-indulgent at times'' statements) positively predict efficient decision-making. Meanwhile, items that are associated with conscientiousness and reliability are either negatively associated with efficiency or weakly positively related (i.e., agreeing with the statement ``I have iron self-discipline''). This might suggest that while a degree of planning and self-regulation improves decision-making, rigid or overly routinized behavior may hinder it. Efficient decision-making, in this sense, requires striking a balance between planning and flexibility. This argument also has somewhat intuitive economic underpinnings: if you insist on purchasing the specific product you decided on beforehand, no matter the price of other closely related products, then you are not making the most efficient decision possible. 

Meanwhile, the SGL manages the objectives of group-level and individual sparsity by selecting a moderate number of predictors (37), sitting between the Lasso (24) and the GL (60). Furthermore, it returns the highest $R^2$ (0.113) and lowest mean-squared error (MSE), while also allowing us to rank the relative importance of each group of variables. This facilitates a more direct point of comparison to the PCR regressions from the previous section. We provide this detailed breakdown of the blocks of variables in the SGL and their relative importance in \cref{tab:group_lasso}.

\begin{table}[!htb]
    \centering
    \scriptsize
    \setlength\tabcolsep{0pt}
    \input{results/tables/lasso/sparse_group_importance.tex}
    \caption{Relative importance of groups from the SGL with $\widehat{AEI}$ as the dependent variable. The groups are ranked according to the percentage of the explained variation they contribute to. We provide the full overview of the individual coefficients within each group in the third column of \cref{tab:full_lasso}.}
    \label{tab:group_lasso}
\end{table}

There appears to be a clear hierarchy in the determinants of decision-making ability. From \cref{tab:group_lasso}, we can immediately gather that sociodemographics dominate in terms of explained variation. The two most important groups: Age (16.6\%) and Household (16.2\%), account for nearly a third of the total explanatory power in the model. Employment (10.8\%) and Gender (7.8\%) also contribute substantially, collectively bringing the demographic share beyond 50\%. Notably, education, income, and geographical location contribute next to nothing in terms of predictive power in the SGL (less than 3\%). This result is in stark contrast to \textcite{echenique_money_2011}, \textcite{choi_who_2014}, and \textcite{dean_measuring_2016}, who all find that both income and education are important predictors of their respective measures of departures from the utility-maximization hypothesis. A possible explanation for this discrepancy is our earlier finding: that it is concern or worry about your financial situation that determines your decision-making ability, not your actual income or retirement savings.\footnote{This argument is also substantiated by our earlier results, where income is a significant predictor (column 1 of \cref{tab:regressions}) until we include the financial concern component.}

In comparison, the two statistically significant components from earlier, Perceived Stress \& Anxiety (10.9\%) and Financial Concerns (7.7\%), only explain 18.6\% of the variation in decision-making quality. Interestingly, many of the items that were not included in any of the principal components (the ``Other behavioral statements'' group) account for 14.8\% of the explained variation. In particular, having planned your shopping ahead of time (``Plan shopping ahead?'') emerges as a significant predictor of decision-making ability within this group. Meanwhile, the components Spontaneity \& Disinhibition and Conscientiousness \& Reliability contribute moderately to predicting decision-making, explaining 6.4\% and 4.9\%, respectively. The remaining three components have negligible effects. 

Overall, the SGL analysis complements the principal components results while adding context to which groups of variables and individual items matter most. Both methods converge on the conclusion that demographics, and age in particular, are the strongest predictors of decision-making quality. Among the behavioral factors, financial concern and worry (as reflected in the self-reported anxiety and concerns over financial control) emerge as the most prominent predictors. These insights offer a nuanced view of consumer behavior: efficient decision-makers in these data are not necessarily those with high incomes or long educations, but rather those who are attentive and concerned enough about their spending to respond effectively and deliberately.

\subsection{A Note on Transitivity}
Inefficiencies and low-quality decision-making manifest in two different ways: direct violations of RP conditions (choosing A over B and then B over A) or indirectly/transitively (choosing A over B, B over C, and C over A). Canonically, the transitivity condition is important, yet the experimental literature has provided evidence of situations in which it does not hold (see e.g., \textcite{tversky_intransitivity_1969}, \textcite{loomes_preference_1989}, and \textcite{loomes_observing_1991}), justifying the considerable attention that the transitivity of preferences has received. Transitivity is closely linked to the symmetry of substitution effects (see \textcite{kihlstrom_demand_1976}) as both are, in essence, non-cycling conditions. As noted in \textcite{mas1995microeconomic}---echoing \textcite{samuelson1947foundations}---symmetry is not a very intuitive idea, and it is hard to see why it should hold without the aid of calculus. In choice environments with many goods, individuals need to consider (or act as if they consider) a very large number of item-level trade-offs simultaneously. If done consciously, this is a complex cognitive requirement---presumably made more so when the number of products being considered is large, as it is in our data. 

To explore this further we exploit the difference between the weak axiom of revealed preference (WARP)\footnote{The Weak Axiom of Revealed Preference: $\mathbf{x_{t}} \ne \mathbf{x_{s}}$ and $\mathbf{x}_{t} R_{0} \mathbf{x}_{s} \Rightarrow NOT \; \mathbf{x}_{s} R_{0} \mathbf{x}_{t}.$ WARP was introduced by \textcite{samuelson_note_1938} as an elementary requirement for rational choice. Relative to GARP, the key difference is that WARP drops transitivity from the rationalizability condition. To be consistent with WARP, the consumer only needs to be consistent in all of their direct pairwise revealed preferences. It is worth noting, however, that WARP and GARP are equivalent in a two-good choice environment \autocite{rose_consistency_1958}. In other words, transitivity has no bite in a two-good choice environment such as that investigated in the experiment in \autocite{choi_who_2014}. In datasets with more than two goods, however, the difference between WARP and GARP provides a means by which we can investigate the quantitative importance of transitivity. See \textcite{gale_note_1960} for an early constructed example of this and \textcite{shafer_revealed_1977}, \textcite{peters_warp_1994} and  \textcite{heufer_generating_2014} for further discussions and extensions.} which does not require transitivity, and  GARP which does.\footnote{To be accurate, GARP allows for multi-valued demand \textit{correspondences} (because it allows ``flat spots'' on indifference curves) whereas the Strong Axiom of Revealed Preference (defined as: $\mathbf{x_{t}} \ne \mathbf{x_{s}}$ and $\mathbf{x}_{t} R_{0} \mathbf{x}_{s} \Rightarrow NOT \; \mathbf{x}_{s} P_{0} \mathbf{x}_{t}$) is necessary and sufficient for efficient utility maxmisation \emph{and} single-value demands. WARP drops transitivity relative to SARP. Relative to GARP, it drops transitivity but imposes single-valued demands. \emph{However}, since in our data, no household is ever observed to face the same budget constraint twice, the distinction between GARP and SARP is, for our purposes, empirically immaterial.} If a household satisfies GARP, then it necessarily satisfies WARP. The converse is not true.  Recall that we use a price-resampling method to deal with missing values. It is therefore possible for a household to pass WARP (but not GARP) for one draw of prices and to fail GARP (and necessarily WARP) for another. We use this property to measure how often transitivity matters, or the incidence of transitivity failures in each resample (denoted $\widehat{\rho} \in [0,1]$), as the proportion of draws where WARP and GARP efficiency measures differ for each household. This proportion provides a simple measure of the extent to which transitivity matters to each household. As $\widehat{\rho}$ approaches one, transitivity matters more, whereas when $\widehat{\rho} = 0$, then transitivity does not matter: violations of RP conditions associated with violations of transitivity are also always associated with direct violations, and transitivity adds nothing.

On average, the rate of violations due to transitivity is modest: 7.7$\%$. In other words, the vast majority of violations of RP conditions do not depend on intransitivities. While this is larger than the  2.39\% of cases that \textcite{cherchye_transitivity_2018} documents, it gives the same general, albeit possibly counter-intuitive message, that despite its apparent restrictiveness and with full regard to its theoretical value, transitivity is not especially empirically material.

\begin{table}[!htb]
    \centering
    \scriptsize
    \input{results/tables/regressions/trans_regression_short_results.tex}
    \caption{Standard OLS regression of $\rho$ (the incidence of transitivity) on sociodemographics and principal components. This table only includes variables that are significant at the 10\% level. The full regression table is available from \cref{tab:transitivity_full}.}
    \label{tab:transitivity_short}
\end{table}

In \cref{tab:transitivity_short} we report the results of estimating the share of transitivity failures, $\widehat{\rho}$, against our usual covariates and the principal components with a standard OLS regression. While some significant associations emerge from \cref{tab:transitivity_short}, the included covariates and principal components explain only a modest proportion of the variation in $\rho$. In terms of the principal components, the interpretation of having financial concerns is largely unchanged from the original specifications, where decision-making quality is the dependent variable. Namely, households that express greater concern about their financial situation tend to exhibit fewer transitivity violations. 

A new insight arises from the Shopping Impulsivity component, which shows a positive association with transitivity failures. This is also a very intuitive economic finding: one would expect that more impulsive shoppers tend to make less stable choices, and thus have less coherent preference orderings that result in more lapses of transitivity. Another novel insight is that female heads of household tend to have fewer failures of transitivity. This is noteworthy, as we previously found that females tended to have lower levels of decision-making efficiency. The same reversed pattern applies to the highest income group ($-0.018$, $p < 0.01$), which our previous results also suggested tended to be associated with lower overall decision-making ability.

\section{Conclusions}\label{sec:conclusions}
Based on a comprehensive panel of Danish consumers, we examined the relationship between the quality of everyday economic decision-making and subjective and objective household circumstances. Decision-making is characterized by significant inefficiency with manifestly negative effects on living standards---on average, the cost is equivalent to almost half of consumers' budgets. There is also substantial heterogeneity in economic decision-making across a diverse set of consumers.

However, our key empirical finding is that, while objective material circumstances are associated with different decision-making quality (gender, the presence of children in the household, age, and income all matter when it comes to making good decisions, for example), subjective emotional states are also important. For example, taking two households with similar incomes, the one that worries more about their finances will tend to make more efficient choices. Thus, rather than impairing judgment, concern seems to motivate more careful economic decision-making. While our analysis is correlational, the fact that worry consistently predicts better decision-making across every specification we estimate is difficult to attribute to chance alone. This insight may suggest that interventions designed to promote better decision-making should focus more explicitly on awareness than on providing further education or higher income.

We interpret the fact that a concept of decision-making quality rooted in economic theory and revealed preference is significantly and plausibly associated with psychometrically well-founded subjective personality measures as supportive of economics beyond the usual ``as-if'' justifications. This connection between positive economics, i.e., nonparametric measures of decision-making, and self-reported personality traits suggests that aspects of the self that are otherwise unobservable play a systematic role in predicting economic outcomes. As such, in complex choice environments, understanding the relationship between decision-making and both objective personal characteristics and the individual's emotional state is important for both theory and policy.

\newpage
\appendix \label{sec:appendix}

\renewcommand{\thetable}{\Alph{subsection}.\arabic{table}}
    \setcounter{table}{0}
\renewcommand{\thefigure}{\Alph{subsection}.\arabic{figure}}
    \setcounter{figure}{0}
\renewcommand{\theHfigure}{\Alph{section}.\arabic{figure}}
\renewcommand{\thesubsection}{\Alph{subsection}}

\section*{Appendix}
\subsection{Variable Translations}
\begin{table}[H]
    \centering
    \scriptsize
    \input{results/tables/pca/variable_translations.tex}
    \caption{Shortened versions of original questions.}
    \label{tab:translations}
\end{table}

\newpage
\subsection{Principal Component Analysis}\label{app:pca}
\setcounter{figure}{0}
\setcounter{table}{0}

\begin{table}[H]
    \centering
    \scriptsize
    \input{results/tables/pca/pca_loadings.tex}
    \caption{Factor loadings from principal component analysis with varimax rotation showing the seven extracted components.}
    \label{tab:pca_loadings}
\end{table}

\newpage
\subsection{Lasso}\label{app:lasso}
\setcounter{figure}{0}
\setcounter{table}{0}
    \centering
    \scriptsize
    \input{results/tables/lasso/lasso_coefficients_comparison}

\newpage
\subsection{The Incidence of Transitivity Failures}
\begin{table}[H]
    \centering
    \scriptsize
    \input{results/tables/regressions/trans_regression_results.tex}
    \caption{Standard OLS regression of $\rho$ (the incidence of transitivity) on sociodemographics and principal components.}
    \label{tab:transitivity_full}
\end{table}

\setcounter{biburlnumpenalty}{9000}
\setcounter{biburlucpenalty}{9000}
\setcounter{biburllcpenalty}{9000}
\makeatletter
\def\UrlBreaks{\do\/\do-\do\_} 
\makeatother

\newpage
\printbibliography

\end{document}

%% file: results/tables/descriptive/transactions_stats.tex
\setlength\tabcolsep{0pt}
\begin{tabular*}{\linewidth}{@{\extracolsep{\fill}}>{\raggedright\arraybackslash}p{4cm}c}
\toprule
Variable & Mean (SD)\\
\midrule
\# of days shopped & 129.44 (68.68)\\
\addlinespace
\# of products purchased & 487.8 (263.05)\\
\addlinespace
\# of transactions & 972.05 (569.06)\\
\addlinespace
Products pr. trip (intensity) & 8.06 (4.17)\\
\addlinespace
Average expenditure pr. trip (DKK) & 181.09 (108.23)\\
\addlinespace
\% of prices missing & 0.98 (0.03)\\
\midrule
Observations (N) & 1664 \\
\midrule
\bottomrule
\end{tabular*}

%% file: results/tables/descriptive/all_survey_stats_long.tex
{\scriptsize
\setlength{\LTleft}{0pt}\setlength{\LTright}{0pt}
\begin{longtable}{@{\extracolsep{\fill}}p{0.8\textwidth}r@{}}
\toprule
Statement & Mean (SD) \\
\midrule
\endfirsthead
\toprule
Statement & Mean (SD) \\
\midrule
\endhead
\midrule
\multicolumn{2}{r}{\textit{Continued on next page}} \\
\endfoot
\midrule
Observations (N) & 1664 \\
\bottomrule
\caption{Descriptive statistics for Tangney's self-control scale, Cohen's perceived stress scale, and the other survey questions.}\label{tab:survey_stats_all}
\endlastfoot
\multicolumn{2}{@{}l}{\textbf{Self-Control Scale}} \\
\quad I eat healthy foods & 3.69 (0.85) \\
\quad I have many healthy habits & 3.74 (0.95) \\
\quad I sometimes drink too much alcohol & 1.84 (1.21) \\
\quad I am good at resisting temptation & 3.25 (0.87) \\
\quad I have a hard time breaking bad habits & 2.78 (0.95) \\
\quad I am lazy & 2.04 (0.91) \\
\quad I often say inappropriate things & 1.95 (0.80) \\
\quad I never allow myself to lose control & 2.75 (1.04) \\
\quad I do certain things that are bad for me, if they are fun & 2.67 (0.93) \\
\quad Getting up in the morning is hard for me & 2.05 (1.19) \\
\quad I have trouble saying no & 2.70 (1.13) \\
\quad I change my mind fairly often & 2.08 (0.77) \\
\quad I blurt out whatever is on my mind & 2.29 (0.91) \\
\quad I refuse things that are bad for me & 3.15 (1.01) \\
\quad I spend too much money & 2.20 (1.02) \\
\quad I keep everything neat & 3.24 (1.06) \\
\quad I am self-indulgent at times & 3.03 (0.92) \\
\quad I wish I had more self-discipline & 2.66 (1.10) \\
\quad I am reliable & 4.28 (0.83) \\
\quad I get carried away by my feelings & 2.90 (0.91) \\
\quad I do many things on the spur of the moment & 2.91 (0.87) \\
\quad I don't keep secrets very well & 1.58 (0.78) \\
\quad I have worked or studied all night at the last minute & 2.61 (1.11) \\
\quad I'm not easily discouraged & 3.50 (1.00) \\
\quad I'd be better off if I stopped to think before acting & 2.15 (0.91) \\
\quad Pleasure and fun sometimes keep me from getting work done & 2.13 (0.90) \\
\quad I have trouble concentrating & 2.16 (1.05) \\
\quad I am able to work effectively toward long-term goals & 3.41 (1.01) \\
\quad Sometimes I can't stop myself from doing something, even if I know it is wrong & 2.57 (0.90) \\
\quad I often act without thinking through all the alternatives & 2.13 (0.85) \\
\quad I lose my temper too easily & 2.17 (1.12) \\
\quad I often interrupt people & 2.25 (0.95) \\
\quad I am always on time & 4.24 (0.98) \\
\quad People can count on me to keep the schedule & 4.27 (0.70) \\
\quad People would describe me as impulsive & 2.59 (0.90) \\
\quad People would say that I have an iron self-discipline & 3.57 (0.81) \\
\addlinespace[.75em]
\multicolumn{2}{@{}l}{\textbf{Stress Scale}} \\
\quad Been upset because of something that happened unexpectedly? & 2.05 (0.99) \\
\quad Felt that you were unable to control the important things in your life? & 2.10 (1.07) \\
\quad Felt nervous and stressed? & 2.06 (1.08) \\
\quad Felt confident about your ability to handle your personal problems & 2.96 (1.36) \\
\quad Felt that things were going your way? & 3.24 (1.08) \\
\quad Found that you could not cope with all the things that you had to do? & 2.55 (1.10) \\
\quad Been able to control irritations in your life? & 3.16 (1.13) \\
\quad Felt that you were on top of things? & 3.59 (1.04) \\
\quad Been angered because of things that were outside of your control & 2.50 (1.04) \\
\quad Felt difficulties were piling up so high that you could not overcome them? & 2.14 (1.07) \\
\addlinespace[.75em]
\multicolumn{2}{@{}l}{\textbf{Other Survey Questions}} \\
\multicolumn{2}{@{}l}{\quad\textit{Shopping behavior}} \\
\quad\quad How often do you plan shopping for more than one day? & 3.92 (1.21) \\
\quad\quad How often do you change plans because there is something more tempting in the shop? & 2.82 (0.95) \\
\quad\quad How often do you buy something that you did not plan to buy? & 3.17 (0.97) \\
\addlinespace[.5em]
\multicolumn{2}{@{}l}{\quad\textit{Financial attitude}} \\
\quad\quad I/we have more debt than I/we should & 1.87 (1.28) \\
\quad\quad I feel I/we should save more than I/we do & 2.49 (1.37) \\
\quad\quad I would like to have better control of my/our expenditure & 2.27 (1.33) \\
\quad\quad I save/have saved for my pension & 3.89 (1.34) \\
\addlinespace[.5em]
\multicolumn{2}{@{}l}{\quad\textit{General well-being}} \\
\quad\quad Do you feel pain or distress? & 2.35 (1.10) \\
\quad\quad Are you anxious or depressed? & 1.66 (0.92) \\
\quad\quad Do you feel stress in your everyday life? & 1.89 (1.01) \\
\end{longtable}}

%% file: results/tables/pca/pca_summary.tex
\setlength\tabcolsep{0pt}
\begin{tabular*}{\linewidth}{@{\extracolsep{\fill}}rlccc}
  \toprule
Component & Component Name & Items & Variance Explained & Cronbach's $\alpha$ \\ 
  \midrule
  1 & Perceived Stress and Anxiety &  11 & 0.13 & 0.90 \\ 
    2 & Spontaneity and Disinhibition &  11 & 0.10 & 0.81 \\ 
    3 & Conscientiousness and Reliability &   6 & 0.06 & 0.68 \\ 
    4 & Poor Health Self-Regulation &   5 & 0.06 & 0.71 \\ 
    5 & Financial Concerns &   4 & 0.06 & 0.81 \\ 
    6 & Perceived Control &   4 & 0.05 & 0.74 \\ 
    7 & Shopping Impulsivity &   4 & 0.04 & 0.53 \\ 
   \bottomrule
\end{tabular*}

%% file: results/tables/descriptive/desc_stats.tex
\setlength\tabcolsep{0pt}
\begin{tabular*}{\linewidth}{@{\extracolsep{\fill}}>{\raggedright\arraybackslash}p{4cm}c}
\toprule
Variable & Mean (SD)\\
\midrule
\addlinespace[.75em]
\multicolumn{2}{l}{\textbf{Household composition}}\\
\hspace{1em}\# of people & 1.94 (0.97)\\
\hspace{1em}\# of kids in ages 0-6 & 0.05 (0.28)\\
\hspace{1em}\# of kids in ages 7-14 & 0.12 (0.42)\\
\hspace{1em}\# of kids in ages 15-20 & 0.13 (0.41)\\
\addlinespace[.75em]
\multicolumn{2}{l}{\textbf{Personal characteristics}}\\
\hspace{1em}Single & 0.36 (0.48)\\
\hspace{1em}Age & 57.05 (13.43)\\
\hspace{1em}Female & 0.77 (0.42)\\
\addlinespace[.75em]
\multicolumn{2}{l}{\textbf{Further education}}\\
\hspace{1em}No (further) education & 0.14 (0.35)\\
\hspace{1em}Vocational education & 0.37 (0.48)\\
\hspace{1em}Short education & 0.16 (0.37)\\
\hspace{1em}Medium education & 0.26 (0.44)\\
\hspace{1em}Long education & 0.07 (0.26)\\
\addlinespace[.75em]
\multicolumn{2}{l}{\textbf{Labor market status}}\\
\hspace{1em}Student & 0.02 (0.15)\\
\hspace{1em}Unemployed & 0.06 (0.24)\\
\hspace{1em}Part time & 0.07 (0.26)\\
\hspace{1em}Full time & 0.41 (0.49)\\
\hspace{1em}Early retirement & 0.14 (0.34)\\
\hspace{1em}Retired & 0.3 (0.46)\\
\addlinespace[.75em]
\multicolumn{2}{l}{\textbf{Household income (in DKK)}}\\
\hspace{1em}- 250K & 0.25 (0.43)\\
\hspace{1em}250K - 500K & 0.38 (0.49)\\
\hspace{1em}500K - 800K & 0.21 (0.41)\\
\hspace{1em}800K - & 0.06 (0.24)\\
\midrule
Observations (N) & 1664 \\
\midrule
\bottomrule
\end{tabular*}

%% file: results/tables/descriptive/aei_stats.tex
\setlength\tabcolsep{0pt}
\begin{tabular*}{0.8\linewidth}{@{\extracolsep{\fill}}lc}
\toprule
\hspace{1em}Minimum & 0.03\\
\hspace{1em}Median & 0.52\\
\hspace{1em}Mean (SD) & 0.51 (0.22)\\
\hspace{1em}Maximum & 1.00\\
\bottomrule
\end{tabular*}

%% file: results/tables/regressions/pca_regression_results.tex
\setlength\tabcolsep{0pt}
\renewcommand{\arraystretch}{1.1}
\begin{tabular*}{\linewidth}{@{\extracolsep{\fill}}l@{\hspace{1em}}r@{\hspace{1em}}r@{\hspace{1em}}r}
\toprule
\textbf{Variable} & \multicolumn{1}{c}{\textbf{Controls Only}} & \multicolumn{1}{c}{\textbf{PC Only}} & \multicolumn{1}{c}{\textbf{Full Model}} \\
\midrule
(Intercept) & \makebox[3em][r]{0.605}$^{***}$ (0.025) & \makebox[3em][r]{0.509}$^{***}$ (0.005) & \makebox[3em][r]{0.589}$^{***}$ (0.025) \\
\midrule
\multicolumn{4}{l}{\textit{Principal Components}} \\[6pt]
PC 1: Perceived Stress and Anxiety &  & \makebox[3em][r]{0.029}$^{***}$ (0.005) & \makebox[3em][r]{0.017}$^{**}$ (0.006) \\[6pt]
PC 2: Spontaneity and Disinhibition &  & \makebox[3em][r]{0.005} (0.006) & \makebox[3em][r]{-0.002} (0.006) \\[6pt]
PC 3: Conscientiousness and Reliability &  & \makebox[3em][r]{-0.005} (0.006) & \makebox[3em][r]{-0.006} (0.006) \\[6pt]
PC 4: Poor Health Self-Regulation &  & \makebox[3em][r]{0.009}$^{\dagger}$ (0.005) & \makebox[3em][r]{0.000} (0.005) \\[6pt]
PC 5: Financial Concerns &  & \makebox[3em][r]{0.036}$^{***}$ (0.005) & \makebox[3em][r]{0.017}$^{**}$ (0.006) \\[6pt]
PC 6: Perceived Control &  & \makebox[3em][r]{-0.008} (0.005) & \makebox[3em][r]{-0.005} (0.005) \\[6pt]
PC 7: Shopping Impulsivity &  & \makebox[3em][r]{-0.007} (0.005) & \makebox[3em][r]{-0.005} (0.005) \\[6pt]
\midrule
\multicolumn{4}{l}{\textit{Control Variables}} \\[6pt]
Female & \makebox[3em][r]{-0.037}$^{**}$ (0.012) &  & \makebox[3em][r]{-0.033}$^{**}$ (0.012) \\[6pt]
Multi-person household & \makebox[3em][r]{-0.016} (0.013) &  & \makebox[3em][r]{-0.016} (0.013) \\[6pt]
Kids 0-6 years & \makebox[3em][r]{0.014} (0.028) &  & \makebox[3em][r]{0.010} (0.028) \\[6pt]
Kids 7-14 years & \makebox[3em][r]{0.067}$^{**}$ (0.022) &  & \makebox[3em][r]{0.066}$^{**}$ (0.022) \\[6pt]
Kids 15-20 years & \makebox[3em][r]{0.048}$^{*}$ (0.019) &  & \makebox[3em][r]{0.045}$^{*}$ (0.019) \\[6pt]
Student & \makebox[3em][r]{0.069}$^{\dagger}$ (0.039) &  & \makebox[3em][r]{0.055} (0.039) \\[6pt]
Unemployed & \makebox[3em][r]{0.032} (0.024) &  & \makebox[3em][r]{0.019} (0.024) \\[6pt]
Part-time employed & \makebox[3em][r]{-0.006} (0.021) &  & \makebox[3em][r]{-0.009} (0.021) \\[6pt]
Early retirement & \makebox[3em][r]{0.014} (0.018) &  & \makebox[3em][r]{0.011} (0.018) \\[6pt]
Retired & \makebox[3em][r]{-0.017} (0.028) &  & \makebox[3em][r]{-0.013} (0.028) \\[6pt]
Age 16-34 & \makebox[3em][r]{0.063}$^{*}$ (0.027) &  & \makebox[3em][r]{0.066}$^{*}$ (0.027) \\[6pt]
Age 50-64 & \makebox[3em][r]{-0.036}$^{*}$ (0.016) &  & \makebox[3em][r]{-0.030}$^{\dagger}$ (0.016) \\[6pt]
Age 65+ & \makebox[3em][r]{-0.079}$^{**}$ (0.029) &  & \makebox[3em][r]{-0.066}$^{*}$ (0.029) \\[6pt]
Vocational education & \makebox[3em][r]{-0.002} (0.016) &  & \makebox[3em][r]{0.000} (0.016) \\[6pt]
Short education & \makebox[3em][r]{-0.007} (0.019) &  & \makebox[3em][r]{-0.005} (0.019) \\[6pt]
Medium education & \makebox[3em][r]{-0.013} (0.017) &  & \makebox[3em][r]{-0.011} (0.018) \\[6pt]
Long education & \makebox[3em][r]{-0.000} (0.025) &  & \makebox[3em][r]{0.001} (0.025) \\[6pt]
Income 250K-500K & \makebox[3em][r]{-0.023}$^{\dagger}$ (0.013) &  & \makebox[3em][r]{-0.017} (0.013) \\[6pt]
Income 500K-800K & \makebox[3em][r]{-0.036}$^{*}$ (0.017) &  & \makebox[3em][r]{-0.026} (0.017) \\[6pt]
Income 800K+ & \makebox[3em][r]{-0.044}$^{\dagger}$ (0.025) &  & \makebox[3em][r]{-0.029} (0.026) \\[6pt]
Urban area & \makebox[3em][r]{-0.007} (0.012) &  & \makebox[3em][r]{-0.007} (0.012) \\[6pt]
Capital region & \makebox[3em][r]{-0.026}$^{\dagger}$ (0.014) &  & \makebox[3em][r]{-0.023}$^{\dagger}$ (0.014) \\[6pt]
\midrule
Observations & \multicolumn{1}{c}{N = 1,664} & \multicolumn{1}{c}{N = 1,664} & \multicolumn{1}{c}{N = 1,664} \\
$R^2$ & \multicolumn{1}{c}{0.092} & \multicolumn{1}{c}{0.045} & \multicolumn{1}{c}{0.102} \\
Adjusted $R^2$ & \multicolumn{1}{c}{0.080} & \multicolumn{1}{c}{0.041} & \multicolumn{1}{c}{0.086} \\
\bottomrule
\multicolumn{4}{l}{\footnotesize{$^{***}p<0.001$; $^{**}p<0.01$; $^{*}p<0.05$; $^{\dagger}p<0.1$}} \\
\end{tabular*}

%% file: results/tables/lasso/model_comparison.tex
\setlength\tabcolsep{0pt}
\renewcommand{\arraystretch}{1.1}
\begin{tabular*}{\linewidth}{@{\extracolsep{\fill}}l@{\hspace{1em}}c@{\hspace{1em}}c@{\hspace{1em}}c@{\hspace{1em}}c}
\toprule
\textbf{Model} & \multicolumn{1}{c}{\textbf{MSE}} & \multicolumn{1}{c}{\textbf{$R^2$}} & \multicolumn{1}{c}{\textbf{Adj.\ $R^2$}} & \multicolumn{1}{c}{\textbf{\# coefficients $\neq 0$}} \\
\midrule
Lasso & \makebox[3em][c]{0.04152} & \makebox[3em][c]{0.108} & \makebox[3em][c]{0.095} & \makebox[3em][c]{24} \\
Group Lasso & \makebox[3em][c]{0.04142} & \makebox[3em][c]{0.110} & \makebox[3em][c]{0.077} & \makebox[3em][c]{60} \\
Sparse Group Lasso ($\alpha$ = 0.95) & \makebox[3em][c]{0.04127} & \makebox[3em][c]{0.113} & \makebox[3em][c]{0.093} & \makebox[3em][c]{37} \\
\bottomrule
\end{tabular*}

%% file: results/tables/lasso/sparse_group_importance.tex
\setlength\tabcolsep{0pt}
\renewcommand{\arraystretch}{1.1}
\begin{tabular*}{\linewidth}{@{\extracolsep{\fill}}l@{\hspace{1em}}c@{\hspace{1em}}c@{\hspace{1em}}c@{\hspace{1em}}c}
\toprule
\textbf{Group} & \textbf{Active/Total} & \textbf{$\|\hat{\delta}_{G}\|_2$} & \textbf{Std.\ norm} & \textbf{\% of total} \\
\midrule
Demographics - Age & \makebox[5em][c]{3/3 (100\%)} & \makebox[3em][c]{0.021} & \makebox[3em][c]{0.012} & \makebox[3em][c]{16.6\%} \\
Demographics - Household & \makebox[5em][c]{4/4 (100\%)} & \makebox[3em][c]{0.020} & \makebox[3em][c]{0.010} & \makebox[3em][c]{16.2\%} \\
Other behavioral statements & \makebox[5em][c]{6/11 (55\%)} & \makebox[3em][c]{0.019} & \makebox[3em][c]{0.006} & \makebox[3em][c]{14.8\%} \\
Perceived Stress and Anxiety & \makebox[5em][c]{4/11 (36\%)} & \makebox[3em][c]{0.014} & \makebox[3em][c]{0.004} & \makebox[3em][c]{10.9\%} \\
Demographics - Employment & \makebox[5em][c]{3/5 (60\%)} & \makebox[3em][c]{0.014} & \makebox[3em][c]{0.006} & \makebox[3em][c]{10.8\%} \\
Demographics - Gender & \makebox[5em][c]{1/1 (100\%)} & \makebox[3em][c]{0.010} & \makebox[3em][c]{0.010} & \makebox[3em][c]{7.8\%} \\
Financial Concerns & \makebox[5em][c]{2/4 (50\%)} & \makebox[3em][c]{0.010} & \makebox[3em][c]{0.005} & \makebox[3em][c]{7.7\%} \\
Spontaneity and Disinhibition & \makebox[5em][c]{2/11 (18\%)} & \makebox[3em][c]{0.008} & \makebox[3em][c]{0.002} & \makebox[3em][c]{6.4\%} \\
Conscientiousness and Reliability & \makebox[5em][c]{4/6 (67\%)} & \makebox[3em][c]{0.006} & \makebox[3em][c]{0.002} & \makebox[3em][c]{4.9\%} \\
Perceived Control & \makebox[5em][c]{2/4 (50\%)} & \makebox[3em][c]{0.002} & \makebox[3em][c]{0.001} & \makebox[3em][c]{1.2\%} \\
Demographics - Education & \makebox[5em][c]{1/4 (25\%)} & \makebox[3em][c]{0.001} & \makebox[3em][c]{0.001} & \makebox[3em][c]{0.9\%} \\
Demographics - Location & \makebox[5em][c]{1/2 (50\%)} & \makebox[3em][c]{0.001} & \makebox[3em][c]{0.001} & \makebox[3em][c]{0.9\%} \\
Shopping Impulsivity & \makebox[5em][c]{2/4 (50\%)} & \makebox[3em][c]{0.001} & \makebox[3em][c]{0.000} & \makebox[3em][c]{0.4\%} \\
Demographics - Income & \makebox[5em][c]{2/3 (67\%)} & \makebox[3em][c]{0.000} & \makebox[3em][c]{0.000} & \makebox[3em][c]{0.3\%} \\
Poor Health Self-Regulation & \makebox[5em][c]{0/5 (0\%)} & \makebox[3em][c]{0.000} & \makebox[3em][c]{0.000} & \makebox[3em][c]{0.0\%} \\
\bottomrule
\end{tabular*}

%% file: results/tables/regressions/trans_regression_short_results.tex
\setlength\tabcolsep{0pt}
\renewcommand{\arraystretch}{1.1}
\begin{tabular*}{\linewidth}{@{\extracolsep{\fill}}l@{\hspace{1em}}r}
\toprule
\textbf{Variable} & \multicolumn{1}{c}{\textbf{Coefficient}} \\
\midrule
(Intercept) & \makebox[3em][r]{0.079}$^{***}$ (0.007) \\
\midrule
\multicolumn{2}{l}{\textit{Principal Components (2 of 7 with $p < 0.1$)}} \\[6pt]
PC 5: Financial Concerns & \makebox[3em][r]{-0.003}$^{*}$ (0.002) \\[6pt]
PC 7: Shopping Impulsivity & \makebox[3em][r]{0.005}$^{**}$ (0.001) \\[6pt]
\midrule
\multicolumn{2}{l}{\textit{Control Variables (5 of 22 with $p < 0.1$)}} \\[6pt]
Female & \makebox[3em][r]{-0.009}$^{*}$ (0.003) \\[6pt]
Age 50-64 & \makebox[3em][r]{0.008}$^{\dagger}$ (0.004) \\[6pt]
Age 65+ & \makebox[3em][r]{0.013}$^{\dagger}$ (0.008) \\[6pt]
Income 800K+ & \makebox[3em][r]{-0.018}$^{**}$ (0.007) \\[6pt]
Urban area & \makebox[3em][r]{0.005}$^{\dagger}$ (0.003) \\[6pt]
\midrule
Observations & N = 1,664 \\
R$^2$ & 0.041 \\
Adjusted R$^2$ & 0.024 \\
F Statistic & 2.43 (df = 29; 1634) $p = < 0.001$ \\
\bottomrule
\multicolumn{2}{l}{\footnotesize{$^{***}p<0.001$; $^{**}p<0.01$; $^{*}p<0.05$; $^{\dagger}p<0.1$}} \\
\end{tabular*}

%% file: results/tables/pca/variable_translations.tex
\setlength\tabcolsep{0pt}
\begin{tabular*}{\linewidth}{@{\extracolsep{\fill}}lr}
  \toprule
Original Question & Shortened Version \\ 
  \midrule
How often do you plan shopping for more than one day? & Plan shopping ahead? \\ 
  How often do you change plans because there is something more tempting in the shop? & Change plans for tempting items? \\ 
  How often do you buy something that you did not plan to buy? & Buy unplanned items? \\ 
  I/we have more debt than I/we should & Have more debt than should \\ 
  I feel I/we should save more than I/we do & Should save more \\ 
  I would like to have better control of my/our expenditure & Want better expenditure control \\ 
  I save/have saved for my pension & Save for pension \\ 
  Do you feel pain or distress? & Feel pain/distress? \\ 
  Are you anxious or depressed? & Feel anxious/depressed? \\ 
  Do you feel stress in your everyday life? & Feel everyday stress? \\ 
  Been upset because of something that happened unexpectedly? & Upset by unexpected events? \\ 
  Felt that you were unable to control the important things in your life? & Unable to control important things? \\ 
  Felt nervous and stressed? & Nervous and stressed? \\ 
  Felt confident about your ability to handle your personal problems & Confident handling problems \\ 
  Felt that things were going your way? & Things going your way? \\ 
  Found that you could not cope with all the things that you had to do? & Cannot cope with all tasks? \\ 
  Been able to control irritations in your life? & Control irritations? \\ 
  Felt that you were on top of things? & On top of things? \\ 
  Been angered because of things that were outside of your control & Angered by things outside control \\ 
  Felt difficulties were piling up so high that you could not overcome them? & Difficulties piling up? \\ 
  Getting up in the morning is hard for me & Morning is hard for me \\ 
  I have trouble concentrating & Trouble concentrating \\ 
  I am able to work effectively toward long-term goals & Work effectively toward goals \\ 
  Sometimes I can't stop myself from doing something, even if I know it is wrong & Can't stop wrong actions \\ 
  I often act without thinking through all the alternatives & Act without thinking alternatives \\ 
  People can count on me to keep the schedule & Keep schedule reliably \\ 
  People would say that I have an iron self-discipline & Have iron self-discipline \\ 
   \bottomrule
\end{tabular*}

%% file: results/tables/pca/pca_loadings.tex
\setlength\tabcolsep{0pt}
\begin{tabular*}{\linewidth}{@{\extracolsep{\fill}}lccccccc}
  \toprule
Variable & PC1 & PC2 & PC3 & PC4 & PC5 & PC6 & PC7 \\ 
  \midrule
  Nervous and stressed? & \cellcolor[gray]{0.8} 0.88 &  0.11 &  -0.06 &  0.03 &  0.13 &  -0.08 &  0.06 \\ 
  Difficulties piling up? & \cellcolor[gray]{0.8} 0.85 &  0.14 &  -0.07 &  0.03 &  0.09 &  -0.11 &  0.04 \\ 
  Unable to control important things? & \cellcolor[gray]{0.8} 0.84 &  0.14 &  -0.06 &  0.04 &  0.14 &  -0.09 &  0.04 \\ 
  Upset by unexpected events? & \cellcolor[gray]{0.8} 0.79 &  0.10 &  -0.04 &  -0.02 &  0.01 &  -0.07 &  0.07 \\ 
  Cannot cope with all tasks? & \cellcolor[gray]{0.8} 0.77 &  0.03 &  -0.07 &  0.07 &  0.10 &  0.13 &  0.08 \\ 
  Feel anxious/depressed? & \cellcolor[gray]{0.8} 0.73 &  0.12 &  -0.14 &  0.20 &  0.16 &  -0.13 &  -0.05 \\ 
  Feel everyday stress? & \cellcolor[gray]{0.8} 0.73 &  0.11 &  -0.05 &  0.11 &  0.23 &  -0.08 &  -0.02 \\ 
  Angered by things outside control & \cellcolor[gray]{0.8} 0.71 &  0.18 &  0.03 &  -0.01 &  0.03 &  0.06 &  0.08 \\ 
  Trouble concentrating & \cellcolor[gray]{0.8} 0.57 &  0.38 &  -0.19 &  0.14 &  0.11 &  -0.02 &  0.08 \\ 
  I'm not easily discouraged & \cellcolor[gray]{0.8} -0.44 &  0.01 &  0.39 &  -0.26 &  -0.03 &  0.33 &  0.07 \\ 
  Feel pain/distress? & \cellcolor[gray]{0.8} 0.42 &  0.04 &  -0.03 &  0.14 &  0.07 &  -0.01 &  0.05 \\ 
  I have trouble saying no &  0.37 &  0.23 &  -0.02 &  0.27 &  0.05 &  0.11 &  0.21 \\ 
  Morning is hard for me &  0.35 &  0.30 &  -0.17 &  0.18 &  0.28 &  0.12 &  -0.01 \\ 
  I blurt out whatever is on my mind &  0.15 & \cellcolor[gray]{0.8} 0.71 &  0.00 &  0.05 &  -0.02 &  -0.03 &  0.03 \\ 
  I often interrupt people &  0.13 & \cellcolor[gray]{0.8} 0.69 &  -0.02 &  0.06 &  -0.03 &  -0.02 &  -0.04 \\ 
  I often say inappropriate things &  0.19 & \cellcolor[gray]{0.8} 0.67 &  -0.05 &  0.16 &  0.04 &  -0.04 &  -0.11 \\ 
  Act without thinking alternatives &  0.09 & \cellcolor[gray]{0.8} 0.58 &  -0.08 &  0.09 &  0.00 &  -0.10 &  0.38 \\ 
  I don't keep secrets very well &  0.14 & \cellcolor[gray]{0.8} 0.58 &  -0.16 &  -0.01 &  0.02 &  -0.02 &  0.08 \\ 
  I lose my temper too easily &  0.25 & \cellcolor[gray]{0.8} 0.56 &  0.04 &  -0.02 &  0.08 &  -0.11 &  -0.01 \\ 
  Pleasure and fun sometimes keep me from getting work done &  0.00 & \cellcolor[gray]{0.8} 0.51 &  -0.11 &  0.06 &  0.14 &  0.12 &  0.36 \\ 
  Can't stop wrong actions &  0.11 & \cellcolor[gray]{0.8} 0.50 &  0.07 &  0.24 &  0.14 &  0.07 &  0.42 \\ 
  I'd be better off if I stopped to think before acting &  0.17 & \cellcolor[gray]{0.8} 0.49 &  -0.12 &  -0.08 &  0.07 &  -0.10 &  0.20 \\ 
  I change my mind fairly often &  0.31 & \cellcolor[gray]{0.8} 0.48 &  -0.16 &  0.19 &  0.03 &  0.05 &  0.16 \\ 
  I have worked or studied all night at the last minute &  0.13 & \cellcolor[gray]{0.8} 0.42 &  -0.22 &  0.06 &  0.24 &  0.19 &  0.17 \\ 
  I do certain things that are bad for me, if they are fun &  -0.02 &  0.39 &  0.18 &  0.24 &  0.15 &  0.13 &  0.20 \\ 
  I sometimes drink too much alcohol &  -0.09 &  0.30 &  -0.09 &  0.03 &  0.07 &  0.02 &  -0.03 \\ 
  Keep schedule reliably &  -0.09 &  -0.12 & \cellcolor[gray]{0.8} 0.74 &  0.06 &  -0.16 &  -0.04 &  0.02 \\ 
  I am always on time &  -0.05 &  -0.14 & \cellcolor[gray]{0.8} 0.72 &  0.09 &  -0.13 &  -0.07 &  0.01 \\ 
  I am reliable &  -0.02 &  -0.14 & \cellcolor[gray]{0.8} 0.65 &  -0.03 &  -0.02 &  0.23 &  0.09 \\ 
  Have iron self-discipline &  -0.11 &  -0.13 & \cellcolor[gray]{0.8} 0.56 &  -0.33 &  -0.04 &  0.09 &  -0.08 \\ 
  Work effectively toward goals &  -0.18 &  -0.06 & \cellcolor[gray]{0.8} 0.52 &  -0.26 &  -0.05 &  0.28 &  -0.07 \\ 
  I keep everything neat &  -0.04 &  -0.11 & \cellcolor[gray]{0.8} 0.47 &  -0.22 &  -0.08 &  0.07 &  -0.04 \\ 
  I never allow myself to lose control &  0.07 &  0.17 &  0.27 &  0.18 &  0.10 &  0.27 &  -0.16 \\ 
  I eat healthy foods &  -0.09 &  -0.07 &  0.09 & \cellcolor[gray]{0.8} -0.72 &  -0.14 &  0.08 &  -0.05 \\ 
  I have many healthy habits &  -0.04 &  -0.10 &  0.07 & \cellcolor[gray]{0.8} -0.67 &  -0.13 &  0.14 &  0.06 \\ 
  I am good at resisting temptation &  -0.16 &  -0.13 &  0.17 & \cellcolor[gray]{0.8} -0.56 &  -0.12 &  0.08 &  -0.35 \\ 
  I have a hard time breaking bad habits &  0.28 &  0.24 &  0.08 & \cellcolor[gray]{0.8} 0.52 &  0.15 &  0.03 &  0.22 \\ 
  I am lazy &  0.14 &  0.40 &  -0.15 & \cellcolor[gray]{0.8} 0.46 &  0.20 &  0.06 &  -0.01 \\ 
  People would describe me as impulsive &  -0.08 &  0.29 &  0.06 &  -0.40 &  0.09 &  -0.16 &  0.33 \\ 
  I wish I had more self-discipline &  0.32 &  0.36 &  -0.10 &  0.38 &  0.25 &  0.03 &  0.27 \\ 
  I refuse things that are bad for me &  -0.12 &  0.06 &  0.30 &  -0.32 &  0.08 &  0.20 &  -0.17 \\ 
  Should save more &  0.21 &  0.11 &  -0.00 &  0.09 & \cellcolor[gray]{0.8} 0.84 &  -0.06 &  0.03 \\ 
  Have more debt than should &  0.18 &  0.07 &  -0.09 &  0.11 & \cellcolor[gray]{0.8} 0.82 &  -0.11 &  0.04 \\ 
  Want better expenditure control &  0.25 &  0.14 &  -0.10 &  0.08 & \cellcolor[gray]{0.8} 0.75 &  -0.08 &  0.13 \\ 
  I spend too much money &  0.12 &  0.34 &  -0.02 &  0.21 & \cellcolor[gray]{0.8} 0.56 &  -0.03 &  0.29 \\ 
  Save for pension &  -0.11 &  0.03 &  0.20 &  -0.11 &  -0.37 &  0.12 &  -0.02 \\ 
  Control irritations? &  -0.01 &  -0.04 &  0.06 &  -0.01 &  -0.08 & \cellcolor[gray]{0.8} 0.78 &  -0.02 \\ 
  On top of things? &  -0.33 &  -0.06 &  0.18 &  -0.12 &  -0.13 & \cellcolor[gray]{0.8} 0.72 &  0.01 \\ 
  Things going your way? &  -0.25 &  -0.01 &  0.12 &  -0.13 &  -0.18 & \cellcolor[gray]{0.8} 0.70 &  0.10 \\ 
  Confident handling problems &  0.19 &  0.01 &  0.08 &  -0.02 &  -0.02 & \cellcolor[gray]{0.8} 0.69 &  -0.04 \\ 
  Buy unplanned items? &  0.11 &  0.06 &  -0.05 &  0.18 &  0.09 &  0.03 & \cellcolor[gray]{0.8} 0.59 \\ 
  Change plans for tempting items? &  0.10 &  0.04 &  -0.05 &  0.01 &  0.02 &  -0.06 & \cellcolor[gray]{0.8} 0.56 \\ 
  I do many things on the spur of the moment &  0.01 &  0.35 &  0.20 &  -0.25 &  0.10 &  0.12 & \cellcolor[gray]{0.8} 0.48 \\ 
  I get carried away by my feelings &  0.30 &  0.38 &  0.09 &  -0.03 &  -0.01 &  -0.06 & \cellcolor[gray]{0.8} 0.43 \\ 
  I am self-indulgent at times &  -0.14 &  0.23 &  0.28 &  0.03 &  0.08 &  0.11 &  0.38 \\ 
  Plan shopping ahead? &  0.01 &  0.09 &  0.15 &  -0.17 &  -0.04 &  0.04 &  -0.23 \\ 
  \midrule
  Variance explained & 0.13 & 0.10 & 0.06 & 0.06 & 0.06 & 0.05 & 0.04 \\ 
  Cumulative variance & 0.13 & 0.22 & 0.28 & 0.34 & 0.39 & 0.44 & 0.49 \\ 
  Cronbach's alpha & 0.90 & 0.81 & 0.68 & 0.71 & 0.81 & 0.74 & 0.53 \\ 
   \bottomrule
\end{tabular*}

%% file: results/tables/lasso/lasso_coefficients_comparison.tex
\setlength\tabcolsep{0pt}
\renewcommand{\arraystretch}{1.1}
\begin{longtable}{@{\extracolsep{\fill}}lrrr}
\toprule
\textbf{Variable} & \textbf{LASSO} & \textbf{GL} & \textbf{SGL} \\
\midrule
\endhead
\midrule
\multicolumn{4}{r}{\footnotesize{Continued on next page}} \\
\endfoot
\bottomrule
\caption{Coefficients of LASSO, Group Lasso (GL), and Sparse Group Lasso (SGL).}\label{tab:full_lasso}\\
\endlastfoot
(Intercept) & \makebox[3.5em][r]{0.5084} & \makebox[3.5em][r]{0.5084} & \makebox[3.5em][r]{0.5084} \\
\midrule
\multicolumn{4}{l}{\textbf{Perceived Stress and Anxiety}} \\
Feel anxious/depressed? & \makebox[3.5em][r]{0.0145} & \makebox[3.5em][r]{0.0085} & \makebox[3.5em][r]{0.0135} \\
Unable to control important things? & \makebox[3.5em][r]{0.0002} & \makebox[3.5em][r]{0.0005} & \makebox[3.5em][r]{0.0006} \\
Feel everyday stress? & \makebox[3.5em][r]{0.0001} & \makebox[3.5em][r]{0.0050} & \makebox[3.5em][r]{0.0017} \\
Feel pain/distress? &  & \makebox[3.5em][r]{0.0002} &  \\
Upset by unexpected events? &  & \makebox[3.5em][r]{0.0027} & \makebox[3.5em][r]{0.0000} \\
Nervous and stressed? &  & \makebox[3.5em][r]{0.0004} &  \\
Cannot cope with all tasks? &  &  &  \\
Angered by things outside control &  &  &  \\
Difficulties piling up? &  &  &  \\
I'm not easily discouraged &  &  &  \\
Trouble concentrating &  & \makebox[3.5em][r]{0.0015} &  \\
\midrule
\multicolumn{4}{l}{\textbf{Spontaneity and Disinhibition}} \\
I often interrupt people & \makebox[3.5em][r]{-0.0074} & \makebox[3.5em][r]{-0.0047} & \makebox[3.5em][r]{-0.0080} \\
I often say inappropriate things &  &  &  \\
I change my mind fairly often &  & \makebox[3.5em][r]{-0.0022} & \makebox[3.5em][r]{-0.0002} \\
I blurt out whatever is on my mind &  & \makebox[3.5em][r]{0.0001} &  \\
I don't keep secrets very well &  & \makebox[3.5em][r]{-0.0004} &  \\
I have worked or studied all night at the last minute &  & \makebox[3.5em][r]{-0.0005} &  \\
I'd be better off if I stopped to think before acting &  &  &  \\
Pleasure and fun sometimes keep me from getting work done &  & \makebox[3.5em][r]{0.0012} &  \\
Can't stop wrong actions &  & \makebox[3.5em][r]{0.0007} &  \\
Act without thinking alternatives &  & \makebox[3.5em][r]{-0.0019} &  \\
I lose my temper too easily &  & \makebox[3.5em][r]{-0.0008} &  \\
\midrule
\multicolumn{4}{l}{\textbf{Conscientiousness and Reliability}} \\
I am always on time & \makebox[3.5em][r]{-0.0047} & \makebox[3.5em][r]{-0.0039} & \makebox[3.5em][r]{-0.0051} \\
Have iron self-discipline & \makebox[3.5em][r]{0.0019} & \makebox[3.5em][r]{0.0031} & \makebox[3.5em][r]{0.0031} \\
Work effectively toward goals & \makebox[3.5em][r]{-0.0001} & \makebox[3.5em][r]{-0.0024} & \makebox[3.5em][r]{-0.0010} \\
I keep everything neat &  &  & \makebox[3.5em][r]{0.0010} \\
I am reliable &  & \makebox[3.5em][r]{-0.0003} &  \\
Keep schedule reliably &  & \makebox[3.5em][r]{-0.0002} &  \\
\midrule
\multicolumn{4}{l}{\textbf{Poor Health Self-Regulation}} \\
I eat healthy foods &  &  &  \\
I have many healthy habits &  &  &  \\
I am good at resisting temptation &  &  &  \\
I have a hard time breaking bad habits &  &  &  \\
I am lazy &  &  &  \\
\midrule
\multicolumn{4}{l}{\textbf{Financial Concerns}} \\
Want better expenditure control & \makebox[3.5em][r]{0.0100} & \makebox[3.5em][r]{0.0075} & \makebox[3.5em][r]{0.0095} \\
Have more debt than should & \makebox[3.5em][r]{0.0000} & \makebox[3.5em][r]{0.0044} & \makebox[3.5em][r]{0.0014} \\
Should save more &  & \makebox[3.5em][r]{0.0006} &  \\
I spend too much money &  &  &  \\
\midrule
\multicolumn{4}{l}{\textbf{Perceived Control}} \\
Things going your way? & \makebox[3.5em][r]{-0.0013} & \makebox[3.5em][r]{-0.0009} & \makebox[3.5em][r]{-0.0015} \\
Confident handling problems &  & \makebox[3.5em][r]{-0.0007} & \makebox[3.5em][r]{-0.0002} \\
Control irritations? &  &  &  \\
On top of things? &  &  &  \\
\midrule
\multicolumn{4}{l}{\textbf{Shopping Impulsivity}} \\
Change plans for tempting items? &  & \makebox[3.5em][r]{-0.0012} &  \\
Buy unplanned items? &  & \makebox[3.5em][r]{0.0033} & \makebox[3.5em][r]{0.0001} \\
I get carried away by my feelings &  & \makebox[3.5em][r]{-0.0007} &  \\
I do many things on the spur of the moment &  & \makebox[3.5em][r]{-0.0008} & \makebox[3.5em][r]{-0.0005} \\
\midrule
\multicolumn{4}{l}{\textbf{Demographics - Gender}} \\
Female & \makebox[3.5em][r]{-0.0086} & \makebox[3.5em][r]{-0.0100} & \makebox[3.5em][r]{-0.0098} \\
\midrule
\multicolumn{4}{l}{\textbf{Demographics - Household}} \\
Kids 7-14 years & \makebox[3.5em][r]{0.0160} & \makebox[3.5em][r]{0.0159} & \makebox[3.5em][r]{0.0171} \\
Kids 15-20 years & \makebox[3.5em][r]{0.0085} & \makebox[3.5em][r]{0.0119} & \makebox[3.5em][r]{0.0098} \\
Multi-person household & \makebox[3.5em][r]{-0.0025} & \makebox[3.5em][r]{-0.0067} & \makebox[3.5em][r]{-0.0044} \\
Kids 0-6 years &  & \makebox[3.5em][r]{0.0050} & \makebox[3.5em][r]{0.0002} \\
\midrule
\multicolumn{4}{l}{\textbf{Demographics - Employment}} \\
Retired & \makebox[3.5em][r]{-0.0067} & \makebox[3.5em][r]{-0.0158} & \makebox[3.5em][r]{-0.0126} \\
Student & \makebox[3.5em][r]{0.0037} & \makebox[3.5em][r]{0.0082} & \makebox[3.5em][r]{0.0049} \\
Unemployed &  & \makebox[3.5em][r]{0.0050} & \makebox[3.5em][r]{0.0012} \\
Part-time employed &  & \makebox[3.5em][r]{-0.0012} &  \\
Early retirement &  & \makebox[3.5em][r]{0.0030} &  \\
\midrule
\multicolumn{4}{l}{\textbf{Demographics - Age}} \\
Age 65+ & \makebox[3.5em][r]{-0.0183} & \makebox[3.5em][r]{-0.0012} & \makebox[3.5em][r]{-0.0122} \\
Age 16-34 & \makebox[3.5em][r]{0.0157} & \makebox[3.5em][r]{0.0142} & \makebox[3.5em][r]{0.0166} \\
Age 50-64 & \makebox[3.5em][r]{-0.0021} & \makebox[3.5em][r]{-0.0006} & \makebox[3.5em][r]{-0.0024} \\
\midrule
\multicolumn{4}{l}{\textbf{Demographics - Education}} \\
Vocational education &  & \makebox[3.5em][r]{0.0004} &  \\
Short education &  & \makebox[3.5em][r]{-0.0001} &  \\
Medium education &  & \makebox[3.5em][r]{-0.0007} & \makebox[3.5em][r]{-0.0012} \\
Long education &  & \makebox[3.5em][r]{0.0003} &  \\
\midrule
\multicolumn{4}{l}{\textbf{Demographics - Income}} \\
Income 250K-500K &  & \makebox[3.5em][r]{-0.0005} & \makebox[3.5em][r]{-0.0002} \\
Income 500K-800K &  & \makebox[3.5em][r]{-0.0005} & \makebox[3.5em][r]{-0.0003} \\
Income 800K+ &  & \makebox[3.5em][r]{-0.0002} &  \\
\midrule
\multicolumn{4}{l}{\textbf{Demographics - Location}} \\
Urban area &  & \makebox[3.5em][r]{-0.0000} &  \\
Capital region &  & \makebox[3.5em][r]{-0.0007} & \makebox[3.5em][r]{-0.0011} \\
\midrule
\multicolumn{4}{l}{\textbf{Other behavioral statements}} \\
Plan shopping ahead? & \makebox[3.5em][r]{0.0120} & \makebox[3.5em][r]{0.0112} & \makebox[3.5em][r]{0.0131} \\
Save for pension & \makebox[3.5em][r]{-0.0089} & \makebox[3.5em][r]{-0.0070} & \makebox[3.5em][r]{-0.0092} \\
Morning is hard for me & \makebox[3.5em][r]{0.0073} & \makebox[3.5em][r]{0.0042} & \makebox[3.5em][r]{0.0072} \\
I am self-indulgent at times & \makebox[3.5em][r]{-0.0044} & \makebox[3.5em][r]{-0.0019} & \makebox[3.5em][r]{-0.0057} \\
I do certain things that are bad for me, if they are fun & \makebox[3.5em][r]{0.0002} & \makebox[3.5em][r]{0.0024} & \makebox[3.5em][r]{0.0016} \\
I sometimes drink too much alcohol &  &  &  \\
I never allow myself to lose control &  &  &  \\
I have trouble saying no &  & \makebox[3.5em][r]{0.0022} & \makebox[3.5em][r]{0.0008} \\
I refuse things that are bad for me &  &  &  \\
I wish I had more self-discipline &  & \makebox[3.5em][r]{0.0001} &  \\
People would describe me as impulsive &  & \makebox[3.5em][r]{0.0010} &  \\
\end{longtable}

%% file: results/tables/regressions/trans_regression_results.tex
\setlength\tabcolsep{0pt}
\renewcommand{\arraystretch}{1.1}
\begin{tabular*}{\linewidth}{@{\extracolsep{\fill}}l@{\hspace{1em}}r}
\toprule
\textbf{Variable} & \multicolumn{1}{c}{\textbf{Coefficient}} \\
\midrule
(Intercept) & \makebox[3em][r]{0.079}$^{***}$ (0.007) \\
\midrule
\multicolumn{2}{l}{\textit{Principal Components}} \\[6pt]
PC 1: Perceived Stress and Anxiety & \makebox[3em][r]{0.000} (0.002) \\[6pt]
PC 2: Spontaneity and Disinhibition & \makebox[3em][r]{-0.002} (0.001) \\[6pt]
PC 3: Conscientiousness and Reliability & \makebox[3em][r]{0.001} (0.001) \\[6pt]
PC 4: Poor Health Self-Regulation & \makebox[3em][r]{-0.001} (0.001) \\[6pt]
PC 5: Financial Concerns & \makebox[3em][r]{-0.003}$^{*}$ (0.002) \\[6pt]
PC 6: Perceived Control & \makebox[3em][r]{-0.001} (0.001) \\[6pt]
PC 7: Shopping Impulsivity & \makebox[3em][r]{0.005}$^{**}$ (0.001) \\[6pt]
\midrule
\multicolumn{2}{l}{\textit{Control Variables}} \\[6pt]
Female & \makebox[3em][r]{-0.009}$^{*}$ (0.003) \\[6pt]
Multi-person household & \makebox[3em][r]{-0.005} (0.003) \\[6pt]
Kids 0-6 years & \makebox[3em][r]{0.006} (0.008) \\[6pt]
Kids 7-14 years & \makebox[3em][r]{0.009} (0.006) \\[6pt]
Kids 15-20 years & \makebox[3em][r]{-0.003} (0.005) \\[6pt]
Student & \makebox[3em][r]{0.007} (0.011) \\[6pt]
Unemployed & \makebox[3em][r]{0.004} (0.006) \\[6pt]
Part-time employed & \makebox[3em][r]{0.008} (0.006) \\[6pt]
Early retirement & \makebox[3em][r]{0.003} (0.005) \\[6pt]
Retired & \makebox[3em][r]{0.000} (0.007) \\[6pt]
Age 16-34 & \makebox[3em][r]{-0.011} (0.007) \\[6pt]
Age 50-64 & \makebox[3em][r]{0.008}$^{\dagger}$ (0.004) \\[6pt]
Age 65+ & \makebox[3em][r]{0.013}$^{\dagger}$ (0.008) \\[6pt]
Vocational education & \makebox[3em][r]{-0.000} (0.004) \\[6pt]
Short education & \makebox[3em][r]{0.002} (0.005) \\[6pt]
Medium education & \makebox[3em][r]{0.002} (0.005) \\[6pt]
Long education & \makebox[3em][r]{-0.001} (0.007) \\[6pt]
Income 250K-500K & \makebox[3em][r]{-0.004} (0.004) \\[6pt]
Income 500K-800K & \makebox[3em][r]{-0.006} (0.005) \\[6pt]
Income 800K+ & \makebox[3em][r]{-0.018}$^{**}$ (0.007) \\[6pt]
Urban area & \makebox[3em][r]{0.005}$^{\dagger}$ (0.003) \\[6pt]
Capital region & \makebox[3em][r]{0.006} (0.004) \\[6pt]
\midrule
Observations & N = 1,664 \\
R$^2$ & 0.041 \\
Adjusted R$^2$ & 0.024 \\
F Statistic & 2.43 (df = 29; 1634) $p = < 0.001$ \\
\bottomrule
\multicolumn{2}{l}{\footnotesize{$^{***}p<0.001$; $^{**}p<0.01$; $^{*}p<0.05$; $^{\dagger}p<0.1$}} \\
\end{tabular*}